\documentclass[12pt]{iopart}
\usepackage[english]{babel}
\usepackage{
amssymb,amsfonts,latexsym}
\usepackage[pdftex]{graphicx}
\usepackage{color}

\usepackage{times}
\usepackage[colorlinks,
citecolor=blue,linkcolor=red,urlcolor=blue]{hyperref}

\usepackage{simplewick}
\usepackage{cancel}

\newtheorem{prop}{Proposition}

\newcommand{\be}{\begin{equation}}
\newcommand{\ee}{\end{equation}}
\newcommand{\bea}{\begin{eqnarray}}
\newcommand{\eea}{\end{eqnarray}}

\begin{document}

\title{Exact formulas for the form factors of local operators in the Lieb-Liniger model}

\author{Lorenzo Piroli and Pasquale Calabrese}
\address{SISSA and INFN, via Bonomea 265, 34136 Trieste, Italy}
\date{\today}

\begin{abstract}
We present exact formulas for the form factors of local operators in the repulsive Lieb-Liniger model at finite size. These are essential ingredients for both numerical and analytical calculations. 
From the theory of Algebraic Bethe Ansatz, it is known that the form factors of local operators satisfy a particular type of recursive relations. 
We show that in some cases these relations can be used directly to derive practical expressions in terms of the determinant 
of a matrix whose dimension scales linearly with the system size. 
Our main results are determinant formulas for the form factors of the operators 
$(\Psi^{\dagger}(0))^2\Psi^2(0)$ and $\Psi^{R}(0)$, for arbitrary integer $R$, where $\Psi$, $\Psi^{\dagger}$ are the usual field operators. 
From these expressions, we also derive the infinite size limit of the form factors of  these local operators in the attractive regime.
\end{abstract}

\section{Introduction}

Exactly solvable models play a unique role in theoretical physics and in particular for many-body systems. 
Integrability makes it 
possible to have an analytical control of the underlying physics which is in general out of reach. 
This gives a remarkable opportunity of sharp theoretical investigations of important aspects of both classical and 
quantum mechanics, as reported in many excellent textbooks \cite{books,takahashi,gaudin, thacker, korepin}.
For a long time exactly solvable models have been the ideal theoretical laboratory where physicists could deepen their understanding of many body physics, develop new mathematical tools and test the limits of approximate methods developed to be applied in the study of more complicated systems.
However, integrability in the past has been often relegated to the realm of formal mathematical physics and thought to have little 
contact with the real physical world. This prejudice has been removed during the last decade mainly thanks to 
the crucial experimental advances in the field of ultra cold atoms, which have made it possible to realize physical systems 
that can be described, with a good approximation, by ideal models (see \cite{review, silva} and references therein). 
One of the prototypical exactly solvable models is the Lieb-Liniger gas introduced in \cite{lieb}. 
This is a one-dimensional model of bosons with point-like interaction, closely related to experimentally realized systems of confined 
bosons in one dimension \cite{kinoshita,YY-chip, fabbri}.
Many other integrable models have been engineered in cold atomic laboratories in the last decade, such as 
the Ising spin chain \cite{ising-exp} and one dimensional Fermi gases \cite{yg-exp,murray-rev}.

Although integrable Hamiltonians can be analytically diagonalized, the computation of most of the  physical and experimentally 
measurable observables remains in general an extremely hard task, because of the difficulty in obtaining manageable 
expressions for the form factors of a local operator, 
which are defined as its matrix elements between eigenstates of the model.

In this work we focus on the exact computation of  form factors of local operators in the Lieb-Liniger gas with a finite number 
of particles in a finite system. 
Despite of more than fifty years of intense investigation following the seminal Bethe ansatz solution of the Lieb-Liniger model \cite{lieb}, 
there are still many physically interesting quantities for which no practical analytical expression is available. 
To the date, only the form factors of the fundamental bosonic field \cite{kojima,caux} and of the density \cite{slavnov_90} 
are known in simple enough forms (in a sense which will be clarified later) to allow the computation of 
the equilibrium correlation function at least numerically \cite{caux,calabrese_0,calabrese_1,panfil}.
For the calculation of expectation values of all other local operators in equilibrium states (which are particular limits of 
form factors), other complementary approaches are usually exploited \cite{gangard, cheianov, mussardo, chou,  pozsgay_1}.

Recently, the importance of disposing of simple, practical expressions for the form factors of local operators has also emerged 
in the study of non-equilibrium dynamics of isolated quantum systems following a quantum quench. 
Indeed, it has been shown \cite{essler} that late times after the quench an integrable systems is locally described by a single 
representative Hamiltonian eigenstate which can be constructed by means of a generalized thermodynamic Bethe ansatz.
This approach has been successfully applied to the Lieb-Liniger model \cite{de_nardis,ga-15}, 
to the XXZ spin-chain \cite{amsterdam, budapest}, to transport problems \cite{de_luca}, 
and to the sine-Gordon field theory \cite{bse-14}.
For finite times the same approach can be used in conjunction with numerical 
techniques \cite{essler, de_nard, de_nardis_2} to provide expectation values of local observables as a single sum of intermediate states, instead of a double 
one of the direct approach \cite{fcc-09,gritsev,mossel,iyer}. 
However, the calculation of measurable observables is limited by the (un-)knowledge of the form factors even for the quench dynamics.
Furthermore, the knowledge of more general form factors greatly helps also in the determination of exact formulas for the 
overlaps between initial state and Bethe states in the quench dynamics which nowadays are known only in very few cases 
\cite{fcc-09,de_nardis,p-13,palacios, brockmann_I,cl-14,pc-14}.

The organization of this paper is as follows. 
In section \ref{lieb_liniger} we introduce the Lieb-Liniger gas and its solution via the (Algebraic) Bethe Ansatz. 
For the sake of clarity we summarize the main results of our work in section \ref{results}, 
while their derivation is presented in section \ref{derivation}. 
In section \ref{attractive_case} we consider the attractive Lieb-Liniger gas and we show how our formulas 
can be simplified in the infinite size limit. 
Our conclusions are presented in section \ref{conclusions}, while technical aspects of our work are reported in the appendices.

\section{The Lieb-Liniger model and the Algebraic Bethe Ansatz}\label{lieb_liniger}
The Lieb-Liniger model describes a system of bosons constrained on a one-dimensional system of length $L$ with periodic boundary conditions. The Hamiltonian written in the second quantization formalism is \cite{lieb}
\begin{equation}
H_{LL}=\int_{0}^{L}\ dx \left(\partial_x\Psi^{\dagger}(x)\partial_x\Psi(x)+c\Psi^{\dagger}(x)\Psi^{\dagger}(x)\Psi(x)\Psi(x)\right)\ ,
\label{hamiltonian}
\end{equation}
where $c$ is the coupling constant (which we now take to be positive, $c>0$). The field operators satisfy canonical commutation relations $[\Psi(x),\Psi^{\dagger}(y)]=\delta(x-y)$.

The eigenstates of the system have a well defined number of particles, which is conserved by the Hamiltonian (\ref{hamiltonian}). 
The model was originally solved using the Coordinate Bethe Ansatz \cite{lieb}, but the natural framework for the computation 
of physical quantities like the scalar product of states, form factors and correlation functions is given by the so called 
Algebraic Bethe Ansatz (ABA) and quantum inverse scattering method \cite{korepin,jimbo,kitanine,gohmann}.


One of the fundamental objects in the ABA method is the monodromy matrix
\begin{equation}
T(\lambda)= \left(\begin{array}{cc}
A(\lambda)&B(\lambda)\\
C(\lambda)&D(\lambda)
\end{array}\right)\ ,
\label{monodromy}
\end{equation}
where $A(\lambda)$, $B(\lambda)$, $C(\lambda)$, $D(\lambda)$ are operators acting on a reference state that we indicate with $|0\rangle$. These operators satisfy a set of non-trivial commutation relations encoded in the Yang-Baxter equations. These, in turn, involve another fundamental object of ABA, namely the $R$-matrix, a $4\times 4$ matrix that in our case reads
\begin{equation}
R(\lambda,\mu)=
\left(\begin{array}{cccc}
f(\mu,\lambda)& & & \\
 &g(\mu,\lambda)&1 &\\
 & 1 & g(\mu,\lambda) & \\
 & & & f(\mu,\lambda)
\end{array}\right)\ ,
\label{r_matrix}
\end{equation}
where empty entries of the matrix are defined to be $0$ and where
\begin{equation}
f(\lambda,\mu)=\frac{\lambda-\mu+ic}{\lambda-\mu}, \ \ \ \ \ g(\lambda,\mu)=\frac{i c}{\lambda -\mu}\ .
\label{functions}
\end{equation}
The Hilbert space is generated by the Bethe states defined as
\begin{equation}
\prod_{j=1}^{N}B(\lambda_j)|0 \rangle\ ,
\label{bethe_states}
\end{equation}
with dual states
\begin{equation}
\langle 0|\prod_{j=1}^{N}C(\lambda_j)\ ,
\end{equation}
while the action on the reference state of the operators $A(\lambda)$, $D(\lambda)$ is given by
\begin{equation}
A(\lambda)|0\rangle=a(\lambda)|0\rangle\ , \qquad D(\lambda)|0\rangle=d(\lambda)|0\rangle.
\label{a_d_operators}
\end{equation}
In the Lieb-Liniger model the functions $a(\lambda)$, $d(\lambda)$ are 
\begin{equation}
a(\lambda)=e^{-i\frac{L}{2}\lambda}\ ,\qquad d(\lambda)=e^{i\frac{L}{2}\lambda}\ .
\label{a_d_functions}
\end{equation}
We further define for later convenience the function
\begin{equation}
r(\lambda)=\frac{a(\lambda)}{d(\lambda)}\ .
\label{r_function}
\end{equation}
The parameters $\{\lambda_j\}_{j=1}^{N}$ are called rapidities and can take arbitrary values. The Hamiltonian (\ref{hamiltonian}) can be related to the transfer matrix defined to be the trace of the monodromy matrix (\ref{monodromy}), $\tau(\lambda)=\mathrm{tr}T(\lambda)$ . This makes it possible to define a $1$-to-$1$ map between the $N$-particle eigenstates of the Hamiltonian (\ref{hamiltonian}) and the states of the form (\ref{bethe_states}) characterized by a set of rapidities $\{\lambda_j\}_{j=1}^{N}$ satisfying the Bethe equations
\begin{equation}
e^{-i\lambda_jL}=\prod_{k=1 \atop k\neq j}^{N}\frac{\lambda_k-\lambda_j+ic}{\lambda_k-\lambda_j-ic}\ .
\label{bethe_equations}
\end{equation}
If a state of the form (\ref{bethe_states}) is characterized by a set of rapidities that satisfy the Bethe equations (\ref{bethe_equations}) we call such a state on-shell, otherwise we call it off-shell. On-shell Bethe states have well defined momentum and energy given respectively by
\begin{equation}
P(\{\lambda_j\})=\sum_{j=1}^{N}\lambda_j\ ,\qquad E(\{\lambda_j\})=\sum_{j=1}^{N}\lambda_j^2\ . 
\label{momentum_energy}
\end{equation}
Finally, it is useful to define rescaled operators
\begin{equation}
\mathcal{B}(\lambda)=\frac{1}{d(\lambda)}B(\lambda)\ ,\qquad \mathcal{C}(\lambda)=\frac{1}{d(\lambda)}C(\lambda)\ ,
\label{b_c_operators}
\end{equation}
where $d(\lambda)$ is given in (\ref{a_d_functions}).

The norm of on-shell Bethe states (\ref{bethe_states}) is given by Gaudin's formula \cite{gaudin, korepin, norms}, which in our notations reads
\begin{equation}
\langle 0|\prod_{j=1}^{N}\mathcal{C}(\lambda_j)\prod_{j=1}^{N}\mathcal{B}(\lambda_j)|0 \rangle=c^{N}\prod_{j<k}\frac{(\lambda_j-\lambda_k)^2+c^2}{(\lambda_j-\lambda_k)^2}\mathrm{det}_{N}\mathcal{N}_{jk}\ ,
\label{norm}
\end{equation}
where
\begin{equation}
\mathcal{N}_{jk}=\delta_{jk}\left(L+\sum_{l=1}^{N}K(\lambda_j,\lambda_l)\right)-K(\lambda_j,\lambda_k)\ ,
\label{gaudin}
\end{equation}
\begin{equation}
K(\lambda,\mu)=\frac{2c}{(\lambda-\mu)^2+c^2}\ .
\label{k_function}
\end{equation}
It is possible to explicitly define the action of the field operators $\Psi(0)$, $\Psi^{\dagger}(0)$ on the Bethe states of the form (\ref{bethe_states}), \cite{korepin}. In particular, one can use the following commutation relations \cite{korepin}
\begin{equation}
[\Psi(0),B(\lambda)]=-i\sqrt{c}A(\lambda)\ , \qquad [C(\lambda),\Psi^{\dagger}(0)]=i\sqrt{c}D(\lambda)\ ,
\label{commutation}
\end{equation}
to derive
\begin{equation}
\Psi(0)\prod_{k=1}^NB(\lambda_k)| 0 \rangle =-i\sqrt{c}\sum_{k=1}^N\Lambda_ka(\lambda_k)\prod_{m=1 \atop m\neq k}^NB(\lambda_m)|0\rangle\ ,
\label{field_action}
\end{equation}
\begin{equation}
\langle 0|\prod_{k=1}^NC(\lambda_k)\Psi^{\dagger}(0)=i\sqrt{c}\sum_{k=1}^N\langle 0|\prod_{m=1 \atop m\neq k}^NC(\lambda_m)\widetilde{\Lambda}_kd(\lambda_k)\ ,
\label{field_action_2}
\end{equation}
where
\begin{equation}
\Lambda_k=\prod_{m=1 \atop m\neq k}^{N}f(\lambda_k,\lambda_m),\qquad \widetilde{\Lambda}_k=\prod_{m=1\atop m\neq k}^{N}f(\lambda_m,\lambda_k)\ .
\label{lambda}
\end{equation}

Note that the connection between the ABA and the Coordinate Bethe Ansatz solutions of the Lieb-Liniger model can be seen explicitly by computing the wave function corresponding to a Bethe state of the form (\ref{bethe_states}). Indeed, using the action of the operator $\Psi(x)$ on Bethe states, and standard techniques in ABA \cite{korepin}, it can be seen that the following is valid
\begin{eqnarray}
\fl \langle 0|\Psi(x_N)\ldots \Psi(x_1)\prod_{j=1}^{N}B(\lambda_j)|0\rangle = &\nonumber \\[0.2cm]
\fl\qquad\quad
 =(-i\sqrt{c})^N\exp\left(-i\frac{L}{2}\sum_{k=1}^N\lambda_k\right)\sum_{\mathcal{P}\in S_N}e^{i \sum_{j=1}^N x_j \lambda_{\mathcal{P}_{j}}}\prod_{j<k}\left(1-\frac{i c\, {\rm sgn}(x_k-x_j)}{\lambda_{\mathcal{P}_k}-\lambda_{\mathcal{P}_j}}\right) .
\end{eqnarray}

Consider now the following class of form factors of local operators
\begin{equation}
\langle 0|\prod_{j=1}^{N}\mathcal{C}(\mu_j)(\Psi^{\dagger}(0))^h\Psi^{k}(0)\prod_{j=1}^{M}\mathcal{B}(\lambda_j)|0\rangle\ .
\label{form_factors}
\end{equation}
Note that the above expression is equal to zero unless $N-h=M-k$. The form factor in (\ref{form_factors}) can in principle be computed either by repeated action of the fields (\ref{field_action}), (\ref{field_action_2}) on Bethe states or through integration of wave-functions in the framework of Coordinate Bethe Ansatz. However, using these methods one arrives at formal expressions which in general involve sums of $\sim N!M!$ terms. These expressions are too complicated for both numerical and analytical calculations, there existing no available procedure to simplify them.

In general, the derivation of practical formulas for the form factors of local operators like those in (\ref{form_factors}) for arbitrary $N$ is indeed a very difficult task. In this paper we address this problem and our main results are summarized in the next section.
The form factors of local operators in the Lieb-Liniger model have been also studied in the recent papers \cite{pozsgay_1, pozsgay_2, kormos_2, shashi, de_nardis_3}. In Ref. \cite{pozsgay_3} techniques similar to those exploited in the present work were used to study form factors of local operators in  systems solvable by Nested Bethe Ansatz.

\section{Summary of our results}\label{results}
In this section we summarize the main results of our work. 
These are represented by the determinant formulas (\ref{rho_squared}), (\ref{general_ff}), (\ref{different_momentum}), (\ref{equal_momentum}) 
for the operators $(\Psi^{\dagger}(0))^2\Psi^{2}(0)$ and $\Psi^{R}(0)$ (with $R$ arbitrary integer). 

\subsection{Form factor of $(\Psi^{\dagger}(0))^2\Psi^2(0)$}
The first result is a practical expression for the form factor of the operator $(\Psi^{\dagger}(0))^2\Psi^2(0)$. Let $\{\mu_j\}_{j=1}^{N}$, $\{\lambda_j\}_{j=1}^{N}$ be two sets of rapidities satisfying the Bethe equations (\ref{bethe_equations}) such that $\mu_j\neq \lambda_k$, $\forall j,k =1,\ldots ,N$. 
Then, our result reads
\begin{eqnarray}
\fl\langle 0 | \prod_{j=1}^{N}\mathcal{C}(\mu_j)(\Psi^{\dagger}(0))^2\Psi^2(0)\prod_{j=1}^{N}\mathcal{B}(\lambda_j)|0 \rangle=(-1)^{N}\frac{\mathcal{J}}{6c}\prod_{j,k=1}^{N}\left(\lambda_{jk}+ic\right)&\nonumber \\
\fl \hspace{4cm}\times \prod_{j=1}^{N}\prod_{k=1}^{N}\frac{1}{\lambda_j-\mu_k}\prod_{j=1}^{N}\left(V^{+}_j-V^{-}_j\right)\frac{\mathrm{det}_{N}\left(\delta_{jk}+U_{jk}\right)}{\left(V^{+}_p-V^{-}_p\right)\left(V^{+}_s-V^{-}_s\right)} \ ,&
\label{rho_squared}
\end{eqnarray}
where $\mathcal{B}$, $\mathcal{C}$ are defined in (\ref{b_c_operators}), $\lambda_{jk}=\lambda_{j}-\lambda_{k}$ and
\begin{equation}
\fl V^{\pm}_{j}=\prod_{m=1}^{N}\frac{\mu_m-\lambda_j \pm ic}{\lambda_m-\lambda_j \pm ic}\ ,
\label{vpm}
\end{equation}
\begin{equation}
\fl U_{jk}=\frac{i}{V^+_j-V^{-}_j}\frac{\prod_{m=1}^N(\mu_{m}-\lambda_j)}{\prod_{m=1\atop m\neq j}^{N}(\lambda_m-\lambda_j)}\left[K(\lambda_j,\lambda_k)-K(\lambda_p,\lambda_k)K(\lambda_s,\lambda_j)\right]\ ,
\label{u_matrix}
\end{equation}
\begin{equation}
\fl \mathcal{J}=(P_{\lambda}-P_{\mu})^4-4(P_{\lambda}-P_{\mu})(Q_{\lambda}-Q_{\mu})+3(E_{\lambda}-E_{\mu})^2 ,
\label{jay}
\end{equation}
\begin{equation}
\fl P_{\lambda}=\sum_{j=1}^N\lambda_j\ ,\quad E_{\lambda}=\sum_{j=1}^N\lambda^2_j\ , \quad Q_{\lambda}=\sum_{j=1}^N\lambda^3_j\ ,
\label{charges}
\end{equation}
and analogously for $P_{\mu}$, $E_{\mu}$, $Q_{\mu}$ . 
In the above equation, $K(\lambda,\mu)$ is given in (\ref{k_function}) and the parameters $\lambda_s$ and $\lambda_p$ are two arbitrary complex numbers not necessarily in the set $\{\lambda_j\}_{j=1}^{N}$. 
Note that the form factor (\ref{rho_squared}) does not depend on $\lambda_p$ and $\lambda_s$ as it will be proved in the following section.

A formula for this form factor was already given by B. Pozsgay in \cite{pozsgay_1}, but there it was expressed as the sum of $N(N-1)/2$ determinants of $N\times N$ matrices. In the case of equal sets of rapidities some simplifications occur and Pozsgay's formula remarkably leads to an expression for the thermodynamic limit of the expectation value of $(\Psi^{\dagger}(0))^{2}\Psi^{2}(0)$ (and analogously for $(\Psi^{\dagger}(0))^{K}\Psi^{K}(0)$). However the formulas presented in \cite{pozsgay_1} are not very convenient in the case of different sets of rapidities, and equation (\ref{rho_squared}) is more suitable both for numerical and analytical calculations (see section \ref{discussions} for further discussions). 

\subsection{Form factor of $\Psi^{R}(0)$}
Our second result is the form factor of the operator $\Psi^{R}(0)$ for arbitrary integer $R$. Let $\{\mu_j\}_{j=1}^{N}$, $\{\lambda_j\}_{j=1}^{N+R}$ be two sets of rapidities satisfying the Bethe equations (\ref{bethe_equations}) and such that $\mu_j\neq \lambda_k$, $\forall j =1,\ldots ,N$, $\forall k =1,\ldots ,N+R$. Then, our result reads
\begin{eqnarray}
\fl \langle 0|\prod_{j=1}^{N}\mathcal{C}(\mu_j)\Psi^{R}(0)\prod_{j=1}^{N+R}\mathcal{B}(\lambda_j)|0\rangle =\frac{(i\sqrt{c})^{R}}{c^{2R-1}(R-1)!}(-1)^{N(R-1)}\prod_{j,k=1}^{N+R}\left(\lambda_{jk}+ic\right)&\nonumber\\
\fl\hspace{2cm}\times\prod_{j=1}^{N+R}\prod_{k=1}^{N}\frac{1}{\lambda_j-\mu_k}\prod_{j=1}^{N+R}\left(\widetilde{V}^{+}_{R,j}-\widetilde{V}^{-}_{R,j}\right)\frac{\mathrm{det}_{N+R}\left(\delta_{jk}+\widetilde{U}^{(R)}_{jk}\right)}{\left(\widetilde{V}^{+}_{R,p}-\widetilde{V}^{-}_{R,p}\right)\left(\widetilde{V}^{+}_{R,s}-\widetilde{V}^{-}_{R,s}\right)}\ , &
\label{general_ff}
\end{eqnarray}
where again $\mathcal{B}$, $\mathcal{C}$ are defined in (\ref{b_c_operators}), $\lambda_{jk}=\lambda_{j}-\lambda_{k}$ and
\begin{equation}
\fl \widetilde{V}^{\pm}_{R,j}=\frac{\prod_{m=1}^{N}\mu_m-\lambda_j \pm ic}{\prod_{m=1}^{N+R}\lambda_m-\lambda_j \pm ic}\ ,
\label{wvpm}
\end{equation}
\begin{equation}
\fl \widetilde{U}^{(R)}_{jk}=\frac{i}{\widetilde{V}^+_{R,j}-\widetilde{V}^{-}_{R,j}}\frac{\prod_{m=1}^N(\mu_{m}-\lambda_j)}{\prod_{m=1\atop m\neq j}^{N+R}(\lambda_m-\lambda_j)}\left[K(\lambda_j,\lambda_k)-K(\lambda_p,\lambda_k)K(\lambda_s,\lambda_j)\right]\ .
\label{u_r_matrix}
\end{equation}
In the above expression $K(\lambda,\mu)$ is given in (\ref{k_function}) and $\lambda_p$, $\lambda_s$ are again two arbitrary complex parameters not necessarily in the set $\{\lambda_j\}_{j=1}^{N+R}$. As before, (\ref{general_ff}) does not depend on $\lambda_p$ and $\lambda_s$. 

Note that the form factor of $\Psi(0)$ was first computed in \cite{kojima}, and a simplified expression for it was given in \cite{caux}. 
Using the techniques discussed in the next section, it is not difficult to see that the expression 
presented in \cite{caux} is equivalent to the case $R=1$ of  equation (\ref{general_ff}).

\subsection{Equivalent formulas}
Equation (\ref{rho_squared}) can be cast in different equivalent forms. We give two of them in the following, which are derived in \ref{derivation_equivalent}.

First, suppose that the two Bethe states have different momentum $P_{\lambda}\neq P_{\mu}$. Then (\ref{rho_squared}) can be rewritten as
\begin{eqnarray}
\fl \langle 0|\prod_{j=1}^{N}\mathcal{C}(\mu_j)(\Psi^{\dagger}(0))^2\Psi^2(0)\prod_{j=1}^{N}\mathcal{B}(\lambda_j)|0\rangle \Big|_{P_{\lambda}\neq P_{\mu}}= (-1)^{N+1}\frac{i\mathcal{J}}{6c(P_{\lambda}-P_{\mu})}&\nonumber \\
\fl \hspace{2cm}\times\prod_{j,k=1}^{N}(\lambda_{jk}+ic)\prod_{j=1}^{N}\prod_{k=1}^N\frac{1}{(\lambda_j-\mu_k)}\prod_{j=1}^N(V_j^+-V_j^-)\frac{\mathrm{det}_N\left(\delta_{jk}+U^{(1)}_{jk}\right)}{V_p^{+}-V_{p}^-}\ ,
\label{different_momentum}
\end{eqnarray}
where $\lambda_p$ is again an arbitrary complex parameter, $\lambda_{jk}=\lambda_j-\lambda_k$ and
\begin{equation}
U^{(1)}_{jk}=\frac{i}{V^+_j-V^-_j}\frac{\prod_{m=1}^N(\mu_m-\lambda_j)}{\prod_{m\neq j}^N(\lambda_m-\lambda_j)}[K(\lambda_j,\lambda_k)-K(\lambda_p,\lambda_k)]\ ,\
\label{u_1}
\end{equation}
and where $K(\lambda,\mu)$, $V^{\pm}_j$, $\mathcal{J}$, $P_{\lambda}$ are defined respectively in (\ref{k_function}), (\ref{vpm}), (\ref{jay}), (\ref{charges}). Note that $U^{(1)}_{jk}$ is the same matrix appearing in the form factor of the density operator $\Psi^{\dagger}(0)\Psi(0)$ as first derived in \cite{slavnov_90}.

Consider now the case where the two Bethe states have equal momentum $P_{\lambda}=P_{\mu}$. Then (\ref{rho_squared}) can be rewritten as 
\begin{eqnarray}
\fl \langle 0|\prod_{j=1}^{N}\mathcal{C}(\mu_j)(\Psi^{\dagger}(0))^2\Psi^2(0)\prod_{j=1}^{N}\mathcal{B}(\lambda_j)|0\rangle \Big|_{P_{\lambda}= P_{\mu}}=(-1)^{N+1}\frac{i(E_{\lambda}-E_{\mu})^2}{2Nc} &\nonumber \\
\fl \hspace{2cm} \times \prod_{j,k=1}^{N}(\lambda_{jk}+ic)\prod_{j=1}^{N}\prod_{k=1}^N\frac{1}{(\lambda_j-\mu_k)}\prod_{j=1}^N(V_j^+-V_j^-)\frac{\mathrm{det}_N\left(\delta_{jk}+U^{(2)}_{jk}\right)}{V_p^{+}-V_{p}^-}\ &,
\label{equal_momentum}
\end{eqnarray}
where $\lambda_p$ is an arbitrary complex parameter, $\lambda_{jk}=\lambda_j-\lambda_k$ and
\begin{equation}
\fl U^{(2)}_{jk}=\frac{i}{V^+_j-V^-_j}\frac{\prod_{m=1}^N(\mu_m-\lambda_j)}{\prod_{m\neq j}^N(\lambda_m-\lambda_j)}[K(\lambda_j,\lambda_k)-K(\lambda_p,\lambda_k)]+\frac{i}{V^+_j-V^-_j}K(\lambda_p,\lambda_k)\ ,
\label{u_2}
\end{equation}
and where $K(\lambda,\mu)$, $E_{\lambda}$ are given in (\ref{k_function}), (\ref{charges}) respectively.

We stress again that equations (\ref{rho_squared}), (\ref{general_ff}), (\ref{different_momentum}), (\ref{equal_momentum}) are valid only for sets of rapidities $\{\mu_j\}$, $\{\lambda_j\}$ satisfying the Bethe equations (\ref{bethe_equations}) and such that $\mu_j\neq \lambda_k$ $\forall j,k$. In particular, our formulas cannot be used for the computation of expectation values (corresponding to the case $\{\mu_j\}_{j=1}^{N}=\{\lambda_j\}_{j=1}^{N}$).

\subsection{Numerical checks and discussions}\label{discussions}
All the formulas presented in this work have been numerically checked against exact computations for a small number of particles. 
We remind that the form factors (\ref{form_factors}) can be computed, for small values of $N$, by repeated action of the 
field (\ref{field_action}) on Bethe states. 
We exploited this property and we numerically computed the form factors of $\Psi^{R}(0)$ for $R=1,2,3,4,5$ 
and for a number of particles up to $N+R=9$. This procedure gives the same result of our formula (\ref{general_ff}).
Formulas (\ref{rho_squared}), (\ref{different_momentum}), (\ref{equal_momentum}) for the form factor of $(\Psi^{\dagger}(0))^2\Psi^2(0)$ have been checked numerically for a number of particles up to $N=9$ by comparison with the formulas presented in \cite{pozsgay_1} and we found perfect agreement between the two.

The determinant formulas presented in this work are very convenient for numerical calculations. 
In particular, equation (\ref{equal_momentum}) was recently used in Ref. \cite{de_nardis_2} for a study of the relaxation dynamics of local observables following a quantum quench in the Lieb-Liniger model. In this work, within the quench action method \cite{essler}, the exact time evolution of the normal ordered observable $:\hat{\rho}^2(0):$ (where $\hat{\rho}$ is the density operator) was computed for systems with a number of particles up to $N=96$.

A possible (straightforward) application of equation (\ref{general_ff}) is given by the numerical computation of the class of form 
factors $(\Psi^{\dagger}(0))^K\Psi^{K}(0)$ through a single resolution of the identity, 
both in the repulsive and the attractive case (see section \ref{attractive_case}). 
This can be done very efficiently for example using the so called ABACUS algorithm \cite{calabrese_0, panfil, caux_2, klauser}

The formulas presented in section \ref{results} are also suitable for non-trivial analytical calculations as those performed 
in \cite{shashi, de_nardis_3}. In these works the form factors of the operators $\Psi(0)$, $\Psi^{\dagger}(0)\Psi(0)$ 
were computed in the thermodynamic limit, starting from the corresponding finite size formulas derived in \cite{kojima, caux, slavnov_90}.

Finally, we comment on the fact that equations (\ref{rho_squared}), (\ref{general_ff}), (\ref{different_momentum}), (\ref{equal_momentum}) are valid only for sets of rapidities $\{\mu_j\}$, $\{\lambda_j\}$ with $\mu_j\neq \lambda_k$ $\forall j,k$. Note that even if two sets $\{\mu_j\}$, $\{\lambda_j\}$ correspond to different Bethe states, it might be that $\mu_j=\lambda_k$ for some $j,k$. However, this will not happen in general because of the constrains imposed by the Bethe equations. One case where this happens is given by two different parity invariant sets of rapidities with an odd number of particles $\{\mu_j\}_{j=1}^{2K+1}=\{-\mu_j\}_{j=1}^{2K+1}$, $\{\lambda_j\}_{j=1}^{2K+1}=\{-\lambda_j\}_{j=1}^{2K+1}$ ($K$ being a positive integer). In this case, even if $\{\mu_j\}\neq \{\lambda_j\}$ the value $0$ belongs to both sets of rapidities and our formulas don't apply. This however is a well-known issue true for on-shell form factors (see e.g. \cite{caux, slavnov_90}) and it is not a limitation in practice when computing correlation functions since one can restrict to even numbers of particles as done in \cite{de_nardis_2}.

Form factors for Bethe states corresponding to sets $\{\mu_j\}$, $\{\lambda_j\}$ where $\mu_j=\lambda_k$ for some $j,k$ cannot be obtained as the limit $\lambda_j\to\mu_k$ directly from formulas (\ref{rho_squared}) (\ref{general_ff}). In particular, our formulas cannot be used to compute the expectation value of the operator $(\Psi^{\dagger}(0))^2\Psi^{2}(0)$. This is because all the formulas presented in this work are valid for on-shell rapidities, while one should start from an off-shell expression in order to obtain a meaningful result also in the limit of coinciding rapidities $\lambda_j\to\mu_k$, for some $j,k$.

\section{Properties of the form factors and proof of the determinant formulas}\label{derivation}
In this section we derive the formulas presented above. For the sake of clarity, some of the proofs will be given in the appendices. We begin with some general considerations.

From (\ref{b_c_operators}), (\ref{field_action}) and (\ref{field_action_2}) it is clear that the form factors in (\ref{form_factors}) depend only on the rapidities $\{\mu_j\}_{j=1}^{N}$, $\{\lambda_j\}_{j=1}^{M}$, and on $\{r(\mu_j)\}_{j=1}^{N},\{r(\lambda_j)\}_{j=1}^{M}$,
 i.e. the values of $r(\lambda)$ evaluated at the rapidities. We can thus define the following function
\begin{equation}
\fl \mathcal{G}^{h,k}_{N,M}(\{\mu_j\},\{\lambda_j\},\{r(\mu_j)\},\{r(\lambda_j)\})=\langle 0|\prod_{j=1}^{N}\mathcal{C}(\mu_j)(\Psi^{\dagger}(0))^h\Psi^{k}(0)\prod_{j=1}^{M}\mathcal{B}(\lambda_j)|0\rangle\ .
\label{g_function}
\end{equation}

In the ABA approach $r(\lambda)$ can be considered as a functional parameter. Indeed one could expand the form factor (\ref{g_function}) using relations (\ref{field_action}) and (\ref{field_action_2}) to obtain a formal expression where $r(\lambda)$ is kept as a functional variable. The Lieb-Liniger model results for the form factors are recovered by making a specific choice of the function $r(\lambda)$. In particular, we can choose $r(\lambda)$ in such a way that
\begin{equation}
r(\lambda_j)=e^{-iL\lambda_j}\ ,\qquad r(\mu_j)=e^{-iL\mu_j}\ ,
\label{definition_1}
\end{equation}
consistently with (\ref{a_d_functions}). However, one can make also another choice for the function $r(\lambda)$ which is as follows. First, define the following function of $x$, depending on the parameters $\{\lambda_k\}_{k=1}^{N}$
\begin{equation}
\vartheta_N(\{\lambda_k\};x)= -\prod_{k=1}^{N}\frac{\lambda_k-x+ic}{\lambda_k-x-ic}.
\label{theta_function}
\end{equation}
Then, one can choose the functional parameter $r(\lambda)$ in such a way that
\begin{eqnarray}
r(\lambda_j)=\vartheta_N(\{\lambda_k\};\lambda_j)=\prod_{k=1\atop k\neq j}^{N}\frac{\lambda_k-\lambda_j+ic}{\lambda_k-\lambda_j-ic}\ ,& \label{definition_2.0} \\
r(\mu_j)=\vartheta_M(\{\mu_k\};\mu_j)=\prod_{k=1\atop k\neq j}^{M}\frac{\mu_k-\mu_j+ic}{\mu_k-\mu_j-ic}. &
\label{definition_2}
\end{eqnarray}
Of course, the two definitions (\ref{definition_1}) and (\ref{definition_2.0}), (\ref{definition_2}) are equivalent if the sets $\{\lambda_j\}$, $\{\mu_j\}$ satisfy the Bethe equations (\ref{bethe_equations}), but for arbitrary rapidities they are not. 

Suppose we choose (\ref{definition_2.0}) and (\ref{definition_2}). Then the expression for the form factor becomes a rational function of the rapidities only (that is, no dependence on the functional parameter $r(\lambda)$ remaining). We can then define
\begin{eqnarray}
\fl\mathcal{F}^{h,k}_{N,M}(\{\mu_k\},\{\lambda_k\})=\mathcal{G}^{h,k}_{N,M}(\{\mu_k\},\{\lambda_k\}, \{r(\mu_j)\}, \{r(\lambda_j)\})\Big|_{\{r(\mu_j)\}=\{\vartheta_M(\{\mu_k\},\ \mu_j)\} \atop \{r(\lambda_j)\}=\{\vartheta_N(\{\lambda_k\},\ \lambda_j)\}}\ ,&
\label{on_shell_ff}
\end{eqnarray}
where $\mathcal{G}^{h,k}_{N,M}$ is give in (\ref{g_function}). We henceforth focus on the function $\mathcal{F}^{h,k}_{N,M}$. We stress again that this function is defined for arbitrary values of the rapidities, even though it is physically relevant only for those satisfying the Bethe equations (\ref{bethe_equations}).

We are now ready to present the main ingredient in our derivation of the formulas presented in section \ref{results}. It is given by the following proposition regarding some fundamental properties of the form factors of local operators \cite{korepin, pozsgay_1, pozsgay_2, kormos_2, izergin, pozsgay_3}.
\begin{prop}\label{properties}
Consider the function $\mathcal{F}^{h,k}_{N,M}$ defined in (\ref{on_shell_ff}) (with $0\leq h\leq N$, $0\leq k\leq M$). Then the following properties hold
\begin{enumerate}
\item 
\begin{equation}
\mathcal{F}^{h,k}_{h,k}(\{\mu_j\}_{j=1}^{h},\{\lambda_j\}_{j=1}^{k})=(-i)^{k-h}(\sqrt{c})^{k+h}h!k!\ ;
\label{property_1}
\end{equation}
\item consider $\mu_m\in\{\mu_{j}\}_{j=1}^{N}$; then the asymptotic behavior of $\mathcal{F}^{h,k}_{N,M}$ as a function of $\mu_m$ is given as follows
\begin{equation}
\fl \lim_{\mu_m\to\infty}\mathcal{F}^{h,k}_{N,M}(\{\mu_j\},\{\lambda_j\})=\left\{
	\begin{array}{cc}
	\fl 0 \ , & h=0\ , \\
	(i\sqrt{c})h \mathcal{F}^{h-1,k}_{N-1,M}(\{\mu_j\}_{j\neq m},\{\lambda_j	\})\ , & h> 0\ ; 
	\end{array}
	\right.
\end{equation}
\item consider $\mathcal{F}^{h,k}_{N,M}(\{\mu_j\},\{\lambda_j\})$ as a function of $\mu_m\in\{\mu_j\}_{j=1}^{N}$. Then it is a rational function and its only singularities are first order poles at $\mu_m=\lambda_j$, $j=1,\ldots M$;
\item the residues of the form factors are given by the following recursive relations
\begin{eqnarray}
\fl\mathcal{F}^{h,k}_{N,M}(\{\mu_j\}_{j=1}^{N},\{\lambda_j\}_{j=1}^{M})\Big|_{\mu_m\to\lambda_k}\sim g(\mu_m,\lambda_k)&\nonumber\\
\fl \times\left[\prod_{j=1\atop j\neq m}^{N}f(\mu_j,\mu_m)\prod_{j=1\atop j\neq k}^{M}f(\lambda_k,\lambda_j)-\prod_{j=1\atop j\neq k}^{M}f(\lambda_j,\lambda_k)\prod_{j=1\atop j\neq m}^{N}f(\mu_m,\mu_j)\right]&\nonumber\\
\hspace{6cm}\mathcal{F}^{h,k}_{N-1,M-1}\left(\{\mu_j\}_{j\neq m},\{\lambda_j\}_{j\neq k}\right)\, .&
\label{recursive}
\end{eqnarray}
\end{enumerate}
\end{prop}
Properties $1-4$ are well known in the theory of Algebraic Bethe Ansatz. For completeness, we present their derivation in \ref{proof_properties}.

Note that the recursive relation (\ref{recursive}), first derived in \cite{izergin}, was also discussed in the recent papers \cite{pozsgay_2, kormos_2, pozsgay_3}. In particular, in \cite{pozsgay_2, kormos_2} its connection with recursive relations appearing in integrable quantum field theory were investigated.

Our strategy for deriving the formulas presented in the previous section is very simple. As a first step we show that properties $1-4$ of Prop. \ref{properties} uniquely determine the on-shell form factors (\ref{on_shell_ff}). As a second step, we prove that the determinant formulas of section \ref{results} satisfy these properties. Note that a similar approach was used in Ref. \cite{pozsgay_3} for the study of local operators in nested Bethe Ansatz systems.

The first step of our derivation is given by the following proposition.
\begin{prop}\label{uniqueness}
Let $h, k$ be two fixed integers, with $h,k\geq 0$, and consider a family of functions $\{\mathcal{H}_{N,M}=\mathcal{H}_{N,M}(\{\mu_j\}_{j=1}^{N},\{\lambda_j\}_{j=1}^{M})\}$ depending on two integers $N$, $M$, with $N-h=M-k$. Suppose that $\mathcal{H}_{N,M}$ satisfies properties $1-4$ of Prop. \ref{properties} for every $N\geq h$, $M\geq k$. To be more precise, suppose that, for the given value of $h$, $k$ the following hold
\begin{enumerate}
\item 
\begin{equation}
\mathcal{H}_{h,k}(\{\mu_j\}_{j=1}^{h},\{\lambda_j\}_{j=1}^{k})=(-i)^{k-h}(\sqrt{c})^{k+h}h!k!\ ;
\end{equation}
\item 
\begin{equation}
\fl \lim_{\mu_m\to\infty}\mathcal{H}_{N,M}(\{\mu_j\},\{\lambda_j\})=\left\{
	\begin{array}{cc}
	\fl 0 \ , & h=0\ , \\
	(i\sqrt{c})h \mathcal{F}^{h-1,k}_{N-1,M}(\{\mu_j\}_{j\neq m},\{\lambda_j	\})\ , & h> 0\ ; 
	\end{array}
	\right.
\end{equation}
\item $\mathcal{H}_{N,M}(\{\mu_j\},\{\lambda_j\})$ as a function of $\mu_m\in\{\mu_j\}_{j=1}^{N}$ is a rational function and its only singularities are first order poles at $\mu_m=\lambda_j$, $j=1,\ldots M$;
\item the residues are given by the following recursive relations

\begin{eqnarray}
\fl \mathcal{H}_{N,M}(\{\mu_j\}_{j=1}^{N},\{\lambda_j\}_{j=1}^{M})\Big|_{\mu_m\to\lambda_k}\sim g(\mu_m,\lambda_k)& \nonumber\\
\fl \hspace{2cm} \times\left[\prod_{j=1\atop j\neq m}^{N}f(\mu_j,\mu_m)\prod_{j=1\atop j\neq k}^{M}f(\lambda_k,\lambda_j)-\prod_{j=1\atop j\neq k}^{M}f(\lambda_j,\lambda_k)\prod_{j=1\atop j\neq m}^{N}f(\mu_m,\mu_j)\right]&\nonumber \\
\fl \hspace{6cm}\times\mathcal{H}_{N-1,M-1}\left(\{\mu_j\}_{j\neq m},\{\lambda_j\}_{j\neq k}\right)\ .&
\label{recursive_2}
\end{eqnarray}
\end{enumerate}
Then 
\begin{equation}
\mathcal{H}_{N,M}(\{\mu_j\}_{j=1}^{N},\{\lambda_j\}_{j=1}^{M})=\mathcal{F}^{h,k}_{N,M}(\{\mu_j\}_{j=1}^{N},\{\lambda_j\}_{j=1}^{M})\ .
\label{thesis}
\end{equation}
\end{prop}
We present the proof of this proposition in \ref{proof_uniqueness}. This proof closely follows the one given by Slavnov 
in \cite{slavnov_89} for the scalar product formula between on-shell and off-shell Bethe states.

In order to prove the formulas presented in section \ref{results}, all we need to do is then to show that they indeed satisfy the properties $1-4$ of Prop. \ref{properties}. This is done in the following, where we separately consider the cases of the form factors of $\Psi^{R}(0)$ and of $(\Psi^{\dagger}(0))^2(\Psi^2(0)$).

\subsection{Form factor of $\Psi^{R}(0)$}

Define the rational function $\mathcal{H_{R}}(\{\mu_j\}_{j=1}^{N},\{\lambda_j\}_{j=1}^{N+R})$ as
\begin{eqnarray}
\fl \mathcal{H_{R}}(\{\mu_j\},\{\lambda_j\})=\frac{(i\sqrt{c})^{R}}{c^{2R-1}(R-1)!}(-1)^{N(R-1)}\prod_{j,k=1}^{N+R}\left(\lambda_{jk}+ic\right)&\nonumber\\
\fl\hspace{2cm} \times\prod_{j=1}^{N+R}\prod_{k=1}^{N}\frac{1}{\lambda_j-\mu_k}\prod_{j=1}^{N+R}\left(\widetilde{V}^{+}_{R,j}-\widetilde{V}^{-}_{R,j}\right)\frac{\mathrm{det}_{N+R}\left(\delta_{jk}+\widetilde{U}^{(R)}_{jk}\right)}{\left(\widetilde{V}^{+}_{R,p}-\widetilde{V}^{-}_{R,p}\right)\left(\widetilde{V}^{+}_{R,s}-\widetilde{V}^{-}_{R,s}\right)}\ ,&
\label{h_function}
\end{eqnarray}
where $\widetilde{V}^{\pm}_{R,j}$, $\widetilde{U}_{jk}^{(R)}$ are given in (\ref{wvpm}) and (\ref{u_r_matrix}). Before showing that (\ref{h_function}) gives the correct expression for the form factors of $\Psi^{R}(0)$, hence proving (\ref{general_ff}), we briefly discuss some properties of the function $\mathcal{H}_{R}$. 

The matrix $\delta_{jk}+\widetilde{U}^{(R)}_{jk}$ in (\ref{h_function}) has a similar structure to those appearing in the formulas for the form factors of $\Psi(0)$ and $\Psi^{\dagger}(0)\Psi(0)$ as computed in \cite{kojima, caux, slavnov_90}. When studying the properties of the determinant of such matrices, a set of identities regarding the sum of rational functions are important \cite{slavnov_90, kitanine_2}. For completeness we discuss them in \ref{useful_identities}. 

Using these identities, it is easy to prove for example that $\mathcal{H}_{R}$ does not depend on the parameters $\lambda_p$ and $\lambda_s$. This is done in the following way. Define for convenience
\begin{equation}
\Xi^{(R)}_j=\frac{\prod_{k=1}^{N}(\mu_k-\lambda_j)}{\prod_{k=1\atop k\neq j}^{N+R}(\lambda_k-\lambda_j)}\ , \qquad \Theta^{(R)}_{j}=\widetilde{V}^{+}_{R,j}-\widetilde{V}^{-}_{R,j}\ .
\label{xi}
\end{equation}
For $k=2,\ldots, N+R$ add to the first column of the matrix $\delta_{jk}+\widetilde{U}^{(R)}_{jk}$ column $k$ multiplied by $\Xi^{(R)}_{k}/\Xi^{(R)}_{1}$. From identity (\ref{identity_1}) in \ref{useful_identities} it follows that the first column becomes proportional to $\widetilde{V}_{R,p}^{+}-\widetilde{V}^{-}_{R,p}$. Exploiting the multilinearity of the determinant we get
\begin{equation}
\frac{\mathrm{det}_{N+R}\left(\delta_{jk}+\widetilde{U}^{(R)}_{jk}\right)}{\widetilde{V}^{+}_{R,p}-\widetilde{V}^{-}_{R,p}}=\frac{\mathrm{det}_{N+R}\left(\mathcal{M}^{(R)}_{jk}\right)}{\Xi_{1}^{(R)}}\ ,
\label{aux_a}
\end{equation}
where
\begin{equation}
\mathcal{M}^{(R)}_{jk}=\left\{
	\begin{array}{cc}
	\Xi_{j}^{(R)}\frac{K(\lambda_s,\lambda_j)}{\widetilde{V}^+_{R,j}-\widetilde{V}^{-}_{R,j}}\ , & \mathrm{if\ } k=1\ ,\\
	\delta_{jk}+\widetilde{U}^{(R)}_{jk}\ , &\mathrm{otherwise\ }.
	\end{array}
	\right.
\end{equation}
Now, for $k=2,\ldots, N+R$, add to column $k$ of matrix $\mathcal{M}^{(R)}_{jk}$ column $1$ multiplied by $iK(\lambda_p,\lambda_k)$. Exploiting again the multilinearity of the determinant we get 
\begin{equation}
\mathrm{det}_{N+R}\left(\mathcal{M}^{(R)}_{jk}\right)=\mathrm{det}_{N+R}\left(\widetilde{\mathcal{M}}^{(R)}_{jk}\right)\ ,
\end{equation}
where
\begin{equation}
\widetilde{\mathcal{M}}^{(R)}_{jk}=\left\{
	\begin{array}{cc}
	\Xi_{j}^{(R)}\frac{K(\lambda_s,\lambda_j)}{\widetilde{V}^+_{R,j}-\widetilde{V}^{-}_{R,j}}\ , & \mathrm{if\ } k=1\ ,\\
	\delta_{jk}+\frac{i}{\widetilde{V}^+_{R,j}-\widetilde{V}^{-}_{R,j}}\Xi_{j}^{(R)}K(\lambda_j,\lambda_k)\ , &\mathrm{otherwise\ }.
	\end{array}
	\right.
	\label{aux_b}
\end{equation}
In the final expression (\ref{aux_b}) $\lambda_p$ has disappeared: we conclude that the l.h.s. of (\ref{aux_a}) and thus $\mathcal{H}_R$ in (\ref{h_function}) are independent of the parameter $\lambda_p$. 

To show that $\mathcal{H}_{R}$ does not depend on $\lambda_s$ we proceed as follows. For $j=2,\ldots, N+R$ add to the first row of the matrix $\delta_{jk}+\widetilde{U}^{(R)}_{jk}$ row $j$ multiplied by $\Theta^{(R)}_{j}/\Theta^{(R)}_{1}$. Using again (\ref{identity_1}) of \ref{useful_identities}, we see that the first row becomes proportional to $\widetilde{V}_{R,s}^{+}-\widetilde{V}^{-}_{R,s}$. We can then follow a similar procedure to the one used before to conclude that $\mathcal{H}_{R}$ does not depend on $\lambda_s$ either. In a similar fashion, one can prove that the r.h.s. of (\ref{rho_squared}) is independent of both $\lambda_s$ and $\lambda_p$.

We stress here that with the above procedure we also immediately see that $\mathcal{H}_R$, as a function of the parameter $\mu_m$, does not have poles corresponding to the zeroes of $(\widetilde{V}_{R,p}^{+}-\widetilde{V}_{R,p}^{-})$ and $(\widetilde{V}_{R,s}^{+}-\widetilde{V}_{R,s}^{-})$, since these factors are canceled by the determinant in the numerator in the r.h.s. of Eq. (\ref{h_function}).

We shall now show that $\mathcal{H}_{R}$ satisfies properties $1-4$ of Prop. \ref{properties}. We begin with property $1$ namely that $\mathcal{H}_{R}(\emptyset,\{\lambda_j\}_{j=1}^{R})=(-i\sqrt{c})^RR!$. Using definition (\ref{h_function}) our goal is to prove that
\begin{eqnarray}
\fl \frac{(i\sqrt{c})^{R}}{c^{2R-1}(R-1)!}\prod_{j,k=1}^{R}(\lambda_{jk}+ic)
\frac{\prod_{j=1}^{R}(\widetilde{V}^{+}_{R,j}-\widetilde{V}^{-}_{R,j})   \det_{R}(\delta_{jk}+\widetilde{U}^{(R)}_{jk})}{(\widetilde{V}^{+}_{R,p}-\widetilde{V}^{-}_{R,p})(\widetilde{V}^{+}_{R,s}-\widetilde{V}^{-}_{R,s})}=(-i\sqrt{c})^R R! ,
\label{to_prove}
\end{eqnarray}
where $\widetilde{V}^{\pm}_{R,j}$ and $\widetilde{U}^{(R)}_{jk}$ are given in (\ref{wvpm}), (\ref{u_r_matrix}). This can be seen by induction.

The case $R=1$ is trivial by direct computation. Supposing now equation (\ref{to_prove}) is true for $R-1\geq 1$ we show that it is also true for $R$. Using simple manipulations and the identities discussed in \ref{useful_identities}, it is not difficult to see that the l.h.s. of (\ref{to_prove}) as a function of $\lambda_{R}$ does not have poles, and it is bounded at infinity. Thus, as a consequence of Liouville theorem in complex analysis, it is a constant. In order to determine this constant, we compute the limit $\lambda_R\to\infty$ of the l.h.s. of (\ref{to_prove}).

From (\ref{u_r_matrix}) we see that $\mathrm{det}_{R}(\delta_{jk}+\widetilde{U}^{(R)}_{jk})\propto 1/\lambda^{2}_R$ as $\lambda_R\to\infty$. Expand now the determinant of the matrix $\delta_{jk}+\widetilde{U}^{(R)}_{jk}$ along the last row using Laplace expansion:
\begin{equation}
\fl \mathrm{det}_{R}\left(\delta_{jk}+\widetilde{U}^{(R)}_{jk}\right)=\sum_{k=1}^{R-1}(-1)^{R+k}\widetilde{U}^{(R)}_{R,k}\mathrm{det}_{R-1}(\mathcal{A}_{k})+ (1+\widetilde{U}^{(R)}_{R,R})\mathrm{det}_{R-1}(\mathcal{A}_{R})\ ,
\label{sum}
\end{equation}
where $\mathcal{A}_j$ is the $(N-1)\times (N-1)$ matrix obtained by removing the last row and column $j$ from the matrix $\delta_{jk}+\widetilde{U}^{(R)}_{jk}$. Using definition (\ref{u_r_matrix}) we see that $\widetilde{U}^{(R)}_{R,k}\propto 1/\lambda_{R}^2$ as $\lambda_R\to\infty$. However, for $k=1,\ldots, R-1$ the matrix $\mathcal{A}_k$ contains the column $(\widetilde{U}^{(R)}_{1,R}, \ldots, \widetilde{U}^{(R)}_{R-1,R})^{t}$. Since $\widetilde{U}^{(R)}_{j,R}\propto 1/(\lambda_R)^2$, and $\widetilde{U}^{(R)}_{jk}=\mathcal{O}(1)$ for $j,k,\neq R$ we see that all the terms in (\ref{sum}) are of order $\mathcal{O}(1/\lambda_R)^4$ except for the last one. So
\begin{equation}
\fl \mathrm{det}_{R}\left(\delta_{jk}+\widetilde{U}^{(R)}_{jk}\right)\sim (1+\widetilde{U}^{(R)}_{R,R})\mathrm{det}_{R-1}(\delta_{jk}+\widetilde{U}^{(R-1)}_{jk})\ ,\qquad \mathrm{for\ }\lambda_{R}\to\infty\ .
\end{equation}
From the definition (\ref{wvpm}), (\ref{u_r_matrix}) the following expansions are given by straightforward calculations
\begin{equation}
\fl 1+\widetilde{U}^{(R)}_{RR}=-\frac{1}{\lambda_{R}^{2}}\frac{R(R-1)}{2}c^2+\mathcal{O}(1/\lambda_{R}^4)\ ,
\label{taylor}
\end{equation}
\begin{equation}
\fl \widetilde{V}^{+}_{R,R}-\widetilde{V}^{-}_{R,R}=(-1)^{R}\frac{2i}{c\lambda_{R}^{R-1}}+\mathcal{O}(1/\lambda^{R}_{R})\ ,
\label{taylor_3}
\end{equation}
and also 
\begin{eqnarray}
\fl\prod_{j,k=1}^{R}\left(\lambda_{jk}+ic\right)\frac{\prod_{j=1}^{R-1}\left(\widetilde{V}^{+}_{R,j}-\widetilde{V}^{-}_{R,j}\right)}{\left(\widetilde{V}^{+}_{R,p}-\widetilde{V}^{-}_{R,p}\right)\left(\widetilde{V}^{+}_{R,s}-\widetilde{V}^{-}_{R,s}\right)}=(-1)^{R-1}ic& \nonumber\\
\fl \times\lambda_{R}^{R+1}\prod_{j,k=1}^{R-1}\left(\lambda_{jk}+ic\right)\frac{\prod_{j=1}^{R-1}\left(\widetilde{V}^{+}_{R-1,j}-\widetilde{V}^{-}_{R-1,j}\right)}{\left(\widetilde{V}^{+}_{R-1,p}-\widetilde{V}^{-}_{R-1,p}\right)\left(\widetilde{V}^{+}_{R-1,s}-\widetilde{V}^{-}_{R-1,s}\right)}+\mathcal{O}(\lambda_{R}^{R})\ .&
\label{taylor_2}
\end{eqnarray}
Using (\ref{taylor}), (\ref{taylor_3}), (\ref{taylor_2}) and the inductive hypothesis for $R-1$ we finally have
\begin{eqnarray}
\fl \lim_{\lambda_R\to\infty}\frac{(i\sqrt{c})^{R}}{c^{2R-1}(R-1)!}\prod_{j,k=1}^{R}\left(\lambda_{jk}+ic\right)\prod_{j=1}^{R}\left(\widetilde{V}^{+}_{R,j}-\widetilde{V}^{-}_{R,j}\right)& \nonumber\\
\fl\hspace{2cm} \times \frac{\mathrm{det}_{R}\left(\delta_{jk}+\widetilde{U}^{(R)}_{jk}\right)}{\left(\widetilde{V}^{+}_{R,p}-\widetilde{V}^{-}_{R,p}\right)\left(\widetilde{V}^{+}_{R,s}-\widetilde{V}^{-}_{R,s}\right)}=(-i\sqrt{c})^RR!\ .&
\label{limit_final}
\end{eqnarray}
The first property of Prop. \ref{properties} is thus proven. To see that the second is true, it is sufficient to observe that 
\begin{equation}
\mathcal{H}_{R}(\{\mu_j\},\{\lambda_j\})\propto \frac{1}{\mu^2_m}\to 0\ ,\qquad \mathrm{for\ }\mu_{m}\to\infty\ .
\end{equation}
Furthermore, property $3$ is easily seen to be satisfied using simple manipulations and the identities discussed in \ref{useful_identities}.

Finally, we address property $4$. It is enough to consider the case $\mu_{N}\to \lambda_{N+R}$ because $\mathcal{H}_{R}$ is completely symmetric in the sets $\{\mu_{k}\}$ and $\{\lambda_k\}$ . Define in the following $M=N+R$. In the limit $\mu_{N}\to \lambda_{M}$ the last row of the matrix $\delta_{jk}+\widetilde{U}^{(R)}_{jk}$ becomes $(0,0,\ldots,0,1)$ so it is straightforward to compute
\begin{eqnarray}
\fl \lim_{\mu_N\to\lambda_M} \mathcal{H_{R}}(\{\mu_j\},\{\lambda_j\})= g(\mu_N,\lambda_M)&\nonumber\\
\fl \times\left[\prod_{j=1}^{N-1}f(\mu_j,\mu_N)\prod_{j=1}^{M-1}f(\lambda_M,\lambda_j)-\prod_{j=1}^{M-1}f(\lambda_j,\lambda_M)\prod_{j=1}^{N-1}f(\mu_N,\mu_j)\right] \frac{(i\sqrt{c})^{R}}{c^{2R-1}(R-1)!}&\nonumber\\
\fl \times (-1)^{(N-1)(R-1)}\prod_{j,k=1}^{M-1}\left(\lambda_{jk}+ic\right)\prod_{j=1}^{M-1}\prod_{k=1}^{N-1}\frac{1}{\lambda_j-\mu_k}\prod_{j=1}^{M-1}\left(\widetilde{V}^{+}_{R,j}-\widetilde{V}^{-}_{R,j}\right)&\nonumber\\
\fl \times \frac{1}{\left(\widetilde{V}^{+}_{R,p}-\widetilde{V}^{-}_{R,p}\right)\left(\widetilde{V}^{+}_{R,s}-\widetilde{V}^{-}_{R,s}\right)}\mathrm{det}_{M-1}\left(\delta_{jk}+\widetilde{U}^{(R)}_{jk}\right) .&
\label{property_4}
\end{eqnarray}
where $\widetilde{V}^{\pm}_{R,j}$ and $U^{(R)}_{jk}$ are defined in (\ref{wvpm}), (\ref{u_r_matrix}) for the sets of rapidities $\{\mu_j\}_{j=1}^{N-1}$, $\{\lambda_j\}_{j=1}^{M-1}$. We have thus shown that the function $\mathcal{H}_{R}$ satisfies all the properties of Prop. \ref{properties} and, as a consequence of Prop. \ref{uniqueness}, that equation (\ref{general_ff}) is correct.

\subsection{Form factor of $(\Psi^{\dagger}(0))^2\Psi^2(0)$}

We now show that the determinant formula in (\ref{rho_squared}) satisfies properties $1-4$ of Prop. \ref{properties}.

The form factor of $(\Psi^{\dagger}(0))^2\Psi^2(0)$ corresponds to the case $(h,k)=(2,2)$ of the general formula (\ref{on_shell_ff}). Looking at property $2$ of Prop. \ref{properties} wee see that in order to prove the validity of (\ref{rho_squared}) we also need an expression for the form factor corresponding to the operator $\Psi^{\dagger}(0)\Psi^{2}(0)$, which we give now. Let $\{\mu_j\}_{j=1}^{N}$, $\{\lambda_j\}_{j=1}^{N+1}$ be two sets of rapidities satisfying the Bethe equations (\ref{bethe_equations}). The form factor of the operator $\Psi^{\dagger}(0)\Psi^{2}(0)$ is
\begin{eqnarray}
\fl \langle 0|\prod_{j=1}^{N}\mathcal{C}(\mu_j)\Psi^{\dagger}(0)\Psi^2(0)\prod_{j=1}^{N+1}\mathcal{B}(\lambda_j)|0\rangle =-i\frac{\mathcal{T}}{c\sqrt{c}}\prod_{j,k=1}^{N+1}\left(\lambda_{jk}+ic\right)&\nonumber\\
\fl \hspace{2cm}\times\prod_{j=1}^{N+1}\prod_{k=1}^{N}\frac{1}{\lambda_j-\mu_k}\prod_{j=1}^{N+1}\left(\widetilde{V}^{+}_{1,j}-\widetilde{V}^{-}_{1,j}\right)\frac{\mathrm{det}_{N+1}\left(\delta_{jk}+\widetilde{U}^{(1)}_{jk}\right)}{\left(\widetilde{V}^{+}_{1,p}-\widetilde{V}^{-}_{1,p}\right)\left(\widetilde{V}^{+}_{1,s}-\widetilde{V}^{-}_{1,s}\right)} &,
\label{field_3_formula}
\end{eqnarray}
where as usual $\mathcal{B}$, $\mathcal{C}$ are defined in (\ref{b_c_operators}) and $\lambda_{jk}=\lambda_{j}-\lambda_{k}$. In the above  $\widetilde{V}^{\pm}_{1,j}$, $\widetilde{U}^{(1)}_{jk}$ are given respectively in (\ref{wvpm}) and (\ref{u_r_matrix}) (for $R=1$) while 
\begin{equation}
\mathcal{T}=\frac{1}{2}\left[(P_{\lambda}-P_{\mu})^2-(E_{\lambda}-E_{\mu})\right]\ ,
\end{equation}
where $P_{\lambda}$, $E_{\lambda}$ are defined in (\ref{charges}).

Formula (\ref{field_3_formula}) can be proven using the same strategy of the previous subsection. In particular, one needs to show that (\ref{field_3_formula}) satisfies properties $1-4$ of Prop. \ref{properties}. It is now straightforward to check property $1$, while one proceeds in the same way of the previous subsection to see that properties $3$ and $4$ are valid. We address property $2$. First, note that the following expansions are valid for $\mu_N\to\infty$
\begin{equation}
\fl \frac{\prod_{m=1}^{N}\mu_m-\lambda_j \pm ic}{\prod_{m=1}^{N+1}\lambda_m-\lambda_j \pm ic}\sim \mu_{N} \frac{\prod_{m=1}^{N-1}\mu_m-\lambda_j \pm ic}{\prod_{m=1}^{N+1}\lambda_m-\lambda_j \pm ic}+\mathcal{O}(1)\ ,\qquad \mu_N\to\infty\ ,
\label{eq1}
\end{equation}
\begin{equation}
\fl \prod_{m=1}^{N}(\mu_m-\lambda_j)\sim \mu_N\prod_{m=1}^{N-1}(\mu_m-\lambda_j)+ \mathcal{O}(1)\ ,\qquad \mu_N\to\infty\ ,
\label{eq2}
\end{equation}
\begin{equation}
\fl \mathcal{T}\sim \mu_N^2 + \mathcal{O}(\mu_N)\ ,\qquad \mu_N\to\infty\ .
\label{eq3}
\end{equation}
Using (\ref{eq1}), (\ref{eq2}) and (\ref{eq3}) it is straightforward to compute
\begin{eqnarray}
\fl \lim_{\mu_{N}\to\infty}-i\frac{\mathcal{T}}{c\sqrt{c}}\prod_{j,k=1}^{N+1}\left(\lambda_{jk}+ic\right)\prod_{j=1}^{N+1}\prod_{k=1}^{N}\frac{1}{\lambda_j-\mu_k}\prod_{j=1}^{N+1}\left(\widetilde{V}^{+}_{1,j}-\widetilde{V}^{-}_{1,j}\right)&\nonumber\\
\fl\qquad \qquad \times \frac{\mathrm{det}_{N+1}\left(\delta_{jk}+\widetilde{U}^{(1)}_{jk}\right)}{\left(\widetilde{V}^{+}_{1,p}-\widetilde{V}^{-}_{1,p}\right)\left(\widetilde{V}^{+}_{1,s}-\widetilde{V}^{-}_{1,s}\right)}=i\sqrt{c}\mathcal{H}_{2}(\{\mu_j\}_{j=1}^{N-1},\{\lambda_j\}_{j=1}^{N+1})\ . 
\label{limit_field_3} 
\end{eqnarray}
where $\mathcal{H}_2$ is given in (\ref{h_function}) (for $R=2$). As we proved in the previous subsection
\begin{equation}
\mathcal{H}_{2}(\{\mu_j\}_{j=1}^{N-1},\{\lambda_j\}_{j=1}^{N+1})=\mathcal{F}_{N-1,N+1}^{0,2}(\{\mu_j\}_{j=1}^{N-1},\{\lambda_j\}_{j=1}^{N+1})\ ,
\end{equation}
so we see that property $2$ of Prop \ref{properties} is satisfied by the r.h.s. of (\ref{field_3_formula}) which is thus correct.

We have now all the ingredients to prove formula (\ref{rho_squared}) for the form factor of $(\Psi^{\dagger}(0))^2\Psi^2(0)$: 
one needs to apply the very same procedure just used to prove formula (\ref{field_3_formula}). 
The steps are the same as before and the calculations present no difficulty so we will not present them here.

\section{Form factors in the attractive regime}\label{attractive_case}
In this section we specialise the form factors given in section \ref{results} to the attractive Lieb-Liniger model (i.e. $c<0$) 
when one of the two Bethe states is the ground state, since in this case major simplifications occur and the determinant can be 
carried out explicitly along the lines of \cite{calabrese_1}. 
We start by very briefly introducing the attractive Lieb-Liniger model and we refer to \cite{takahashi, calabrese_1} 
for all the necessary technical details.

We consider the Hamiltonian (\ref{hamiltonian}) in the case $c<0$ and we define for convenience $\overline{c}=-c>0$. With this choice, the Hamiltonian (\ref{hamiltonian}) describes a system of bosons with attractive interaction.
The eigenstates are again in $1$-to-$1$ correspondence with the sets of rapidities satisfying the Bethe equations which now read
\begin{equation}
e^{-i\lambda_jL}=\prod_{k=1\atop k\neq j}^{N}\frac{\lambda_k-\lambda_j-i\overline{c}}{\lambda_k-\lambda_j+i\overline{c}}\ .
\label{bethe_equations_attractive}
\end{equation}
In the attractive regime equations (\ref{bethe_equations_attractive}) admit string solutions corresponding to bound states of bosons \cite{calabrese_1} . Here we focus on the case where all the $N$ particles of the system form a unique bound state, namely a $N$-string. The rapidities forming a $N$-string solution arrange themselves in the following way
\begin{equation}
\lambda_{j}=\lambda+i\frac{\overline{c}}{2}(N+1-2j)+i\delta_{j}\ ,
\label{strings}
\end{equation}
where $\lambda$ is a real parameter (the center of the string), and where $\delta_j$ are small deviations vanishing in the limit where the length $L$ is infinity and the number of particles $N$ is kept fixed. 
The ground state of the system corresponds to a $N$-string centered at $\lambda=0$.

Following \cite{calabrese_1} we now wish to obtain a simplified expression for the form factors presented in section \ref{results} when one of the two Bethe states corresponds to the ground state of the attractive model in the limit of vanishing deviations. 
First, we consider the form factor of $(\Psi^{\dagger}(0))^{2}\Psi^{2}(0)$. Let $\{\mu_{j}\}_{j=1}^{N}$, $\{\lambda_{j}\}_{j=1}^{N}$ be two sets of rapidities both satisfying the Bethe equations (\ref{bethe_equations_attractive}), with $\{\lambda_j\}$ corresponding to the ground state (instead the rapidities $\{\mu_{j}\}_{j=1}^{N}$ do not need to form a $N$-string and correspond to an arbitrary excited state of the system). 
In a finite system the deviations $\delta_j$ will be in general non vanishing and one can plug the rapidities (\ref{strings}) directly in the 
equation (\ref{rho_squared}), since singular terms do not appear. Taking then the limit $\{\delta_j\}\to 0$ the determinant expression gets 
simplified. We report in \ref{attractive_calculations} the calculations. 
The final result for the form factor (with $|\{\lambda_i\}\rangle$ being the ground state) is
\begin{eqnarray}
\fl \langle 0|\prod_{j=1}^{N}\mathcal{C}(\mu_j)(\Psi^{\dagger}(0))^2\Psi^2(0)\prod_{j=1}^{N}\mathcal{B}(\lambda_j)|0\rangle=&\\
\fl 
(-1)^N\frac{c^{2(N-1)}}{6}   \left[P^4_{\mu}-4P_{\mu}Q_{\mu}+3\left(E_{\mu}+\frac{\overline{c}^2}{12}N(N^2-1)\right)^2\right]\frac{N\Gamma^2(N)}{\prod_{m=1}^{N}\left(\mu^2_{m}+\frac{\overline{c}^2}{4}(N-1)^2\right)} .&\nonumber
\label{attractive_rho}
\end{eqnarray}
where $\Gamma$ is the gamma function 
and $P_{\mu}$, $E_{\mu}$, $Q_{\mu}$ are defined in (\ref{charges}).

Consider now the form factor of $\Psi^{R}(0)$. Let $\{\mu_{j}\}_{j=1}^{N}$, $\{\lambda_{j}\}_{j=1}^{N+R}$ be two sets of rapidities satisfying (\ref{bethe_equations_attractive}), with $\{\lambda_{j}\}_{j=1}^{N+R}$ corresponding to the ground state while $\{\mu_{j}\}_{j=1}^{N}$ to an arbitrary excited state of the system. The form factor (\ref{general_ff}) can be simplified in the limit $\{\delta_j\}\to 0$. The calculations are again reported in \ref{attractive_calculations} and the final result for the form factor is
\begin{eqnarray}
\fl \langle 0|\prod_{j=1}^{N}\mathcal{C}(\mu_j)\Psi^R(0)\prod_{j=1}^{N+R}\mathcal{B}(\lambda_j)|0\rangle=&(-1)^{N+R}(i\sqrt{c})^{R}c^{2N}
\nonumber\\\fl & \times 
\frac{N+R}{(R-1)!}\frac{\Gamma^2(N+R)}{\prod_{m=1}^{N}\left(\mu^2_{m}+\frac{\overline{c}^2}{4}(N+R-1)^2\right)}\ .
\label{attractive_field}
\end{eqnarray}

Apart from the per se interest, the attractive Bose gas became recently a standard tool to solve the one dimensional 
Kardar Parisi Zhang equation \cite{KPZ} by means of replicas \cite{plr-10,d-10,cl-11,is-12,gl-12,l-14}.
The form factors provided here are the starting point for the calculation of observables which nowadays are known 
only in the limit of infinite time \cite{d-13,iss-13}. 

\section{Conclusions}\label{conclusions}
In this work we presented exact formulas for the form factors of local operators in the Lieb-Liniger model. 
Our main results are: (i) the determinant formula (\ref{general_ff}) for the form factor of $\Psi^{R}(0)$ with $R$ an arbitrary integer;
(ii) two equivalent formulations for the form factors of $(\Psi^{\dagger}(0))^2\Psi^{2}(0)$, i.e. one determinant formula   
(\ref{rho_squared}) and a set of two determinants (\ref{different_momentum}) and (\ref{equal_momentum}) which are valid respectively for 
different and  equal momenta. 
Along the lines of \cite{calabrese_1}, we also showed how our formulas can be simplified when one is interested in the form factors between the ground state of the attractive model and an arbitrary on-shell Bethe state. Our formulas can be useful for both numerical and analytical calculations, in equilibrium and non-equilibrium settings. 

From the theoretical point of view, equation (\ref{general_ff}) is particularly interesting because it shows that an entire family of local form factors can be expressed as the determinant of a $N\times N$ matrix of simple form, similar to those already known for the form factors of $\Psi(0)$ and $\Psi^{\dagger}(0)\Psi(0)$ \cite{kojima,caux,slavnov_90}. It would be extremely useful to find a similar representation for the form factors of the type $(\Psi^{\dagger}(0))^k\Psi^{k}(0)$ considered in \cite{pozsgay_1}.

Note that in principle the method applied in this work could be used to derive manageable formulas for the general class of form factors (\ref{form_factors}). However, as it should be clear from our derivation in section \ref{derivation}, our approach requires a preliminary guesswork: indeed as a first step one tries to guess, for the desired form factor, an expression that satisfies the properties $1-4$ of Prop. \ref{properties}. The second step, namely the rigorous proof that the properties $1-4$ of Prop. \ref{properties} are indeed satisfied, is straightforward if the initial guess is correct. Of course, the initial guesswork is not blindfold and one is guided by the constrains imposed by Prop. \ref{properties}. 

The starting point for our initial guesswork leading to formulas (\ref{rho_squared}) and (\ref{general_ff}) has been given by the already known determinant formulas for the form factors of $\Psi(0)$ and $\Psi^{\dagger}(0)\Psi(0)$ as first derived in \cite{kojima,caux,slavnov_90}. Our strategy has been to look for ``minimal'' modifications of such formulas that could still satisfy properties $1-4$ of Prop. \ref{properties} for more general form factors. While this approach has turned out to be convenient for the form factors of $\Psi^{R}(0)$ and $(\Psi^{\dagger}(0))^{2}\Psi^2(0)$, we have not yet been able to follow this strategy for the general case of Eq. (\ref{form_factors}).

The Lieb-Liniger model can be obtained as a scaling limit of the XXZ model \cite{gaudin, golzer, seel, pozsgay_1}. 
It is then natural to wonder whether the methods applied in this work could be used also for the $XXZ$ chain. 
In particular, it would be interesting to have a single determinant representation for the form factor 
of $\sigma^{z}_j\sigma^{z}_{j+1}$ (and so giving a simpler representation than the one in \cite{klauser}). 
These issues will be addressed in forthcoming publications.

\section{Acknowledgments}
We acknowledge Jacopo De Nardis for useful discussions and for an independent numerical check of the validity of Eqs. (\ref{different_momentum}), (\ref{equal_momentum}).
PC acknowledges the financial  support by the ERC under  Starting Grant  279391 EDEQS.

\appendix

\section{Proof of Proposition \ref{properties}} \label{proof_properties}
We now present a detailed proof of Prop. \ref{properties}. We begin with property $1$. To do this, we first prove by induction that
\begin{equation}
\Psi^{N}(0)\prod_{j=1}^{N}\mathcal{B}(\lambda)|0\rangle=(-i\sqrt{c})^{N}N!\prod_{j=1}^{N}r(\lambda_j)|0\rangle\ .
\label{a1}
\end{equation}
The case $N=1$ is trivial using (\ref{commutation}). Suppose (\ref{a1}) is true for $N\geq 1$. Then using (\ref{field_action}) we have
\begin{eqnarray}
\fl\Psi^{N+1}(0)\prod_{j=1}^{N+1}\mathcal{B}(\lambda_j)|0\rangle=\Psi^{N}(0)\left[-i\sqrt{c}\sum_{k=1}^{N+1}\Lambda_kr(\lambda_k)\prod_{m=1\atop m\neq k}^{N+1}B(\lambda_m)\right]|0\rangle=(-i\sqrt{c})^{N+1}N!&\nonumber \\
\fl\qquad \times \prod_{m=1}^{N+1}r(\lambda_j)\sum_{k=1}^{N+1}\prod_{m=1\atop m\neq k}^{N+1}f(\lambda_k,\lambda_m)|0\rangle=(-i\sqrt{c})^{N+1}\left(\prod_{j=1}^{N+1}r(\lambda_j)\right)(N+1)!|0\rangle\ ,
\end{eqnarray}
where $\Lambda_k$ is given in (\ref{lambda}) and where we used the inductive hypothesis and the identity
\begin{equation}
\sum_{k=1}^{N+1}\prod_{m=1\atop m\neq k}^{N+1}f(\lambda_k,\lambda_m)=N+1\ .
\end{equation}
Equation (\ref{a1}) is thus proved. Now, for a set of rapidities $\{\lambda_j\}_{}$ satisfying the Bethe equations (\ref{bethe_equations}) we have
\begin{equation}
\prod_{j=1}^{N}r(\lambda_j)=1\ ,
\end{equation}
so (\ref{a1}) becomes
\begin{equation}
\Psi^{N}(0)\prod_{j=1}^{N}\mathcal{B}(\lambda)|0\rangle=(-i\sqrt{c})^{N}N!|0\rangle\ .
\label{a2}
\end{equation}
Analogously one can prove
\begin{equation}
\langle 0|\prod_{j=1}^{N}\mathcal{C}(\mu_j)(\Psi^{\dagger}(0))^{N}=(i\sqrt{c})^{N}N!\langle 0|\ .
\label{a3}
\end{equation}
Using (\ref{a2}), (\ref{a3}) property $1$ of Prop. \ref{properties} is immediately proved.

We now address property $2$. An explicit proof using the Bethe wave functions was given in \cite{pozsgay_3} for the cases $(h,k)=(0,1)$ and $(h,k)=(1,1)$ . For the sake of completeness we give here a proof of the general case. This is based on the properties of the scalar product between two Bethe vectors \cite{korepin, slavnov_89}. It is known from the theory of ABA that this is a rational function of the rapidities and that 
\begin{equation}
\lim_{\mu_m\to\infty}\langle 0|\prod_{j=1}^{N}\mathcal{C}(\mu_j)\prod_{j=1}^{N}\mathcal{B}(\lambda_j)|0\rangle=0\ .
\label{limit_s}
\end{equation}
From (\ref{limit_s}) it is immediately seen by induction and using (\ref{field_action}) that
\begin{equation}
\lim_{\mu_m\to\infty}\langle 0|\prod_{j=1}^{N}\mathcal{C}(\mu_j)\Psi^{k}(0)\prod_{j=1}^{N+k}\mathcal{B}(\lambda_j)|0\rangle =0\ .
\label{a4}
\end{equation}
We now prove, once again by induction, that

\begin{eqnarray}
\fl\lim_{\mu_m\to\infty}\langle 0|\prod_{j=1}^{N+h}\mathcal{C}(\mu_j)(\Psi^{\dagger}(0))^h\Psi^{k}(0)\prod_{j=1}^{N+k}\mathcal{B}(\lambda_j)|0\rangle=&\nonumber\\ \qquad \qquad 
\fl =(i\sqrt{c})h\langle 0|\prod_{j=1\atop j\neq m}^{N+h}\mathcal{C}(\mu_j)(\Psi^{\dagger}(0))^{h-1}\Psi^{k}(0)\prod_{j=1}^{N+k}\mathcal{B}(\lambda_j)|0\rangle \ ,\qquad h\geq 1\ .&
\label{a5}
\end{eqnarray}
The case $h=1$ is immediately seen using (\ref{field_action_2}) and (\ref{a4}). Suppose now (\ref{a5}) is true for $h\geq 1$. Using (\ref{field_action_2}) we have
\begin{eqnarray}
\fl \langle 0|\prod_{j=1}^{N+h+1}\mathcal{C}(\mu_j)(\Psi^{\dagger}(0))^{h+1}\Psi^{k}(0)\prod_{j=1}^{N+k}\mathcal{B}(\lambda_j)|0\rangle=& \nonumber\\\qquad 
\fl =i\sqrt{c}\sum_{t=1}^{N+h+1}\prod_{o=1\atop o\neq t}^{N+h+1}f(\mu_o,\mu_t)\langle 0|\prod_{j=1\atop j\neq t}^{N+h+1}\mathcal{C}(\mu_j)(\Psi^{\dagger}(0))^h\Psi^{k}(0)\prod_{j=1}^{N+k}\mathcal{B}(\lambda_j)|0\rangle .&
\label{a6}
\end{eqnarray}
We take now the limit $\mu_m\to \infty$ using the inductive hypothesis
\begin{eqnarray}
\fl \lim_{\mu_m\to\infty}\langle 0|\prod_{j=1}^{N+h+1}\mathcal{C}(\mu_j)(\Psi^{\dagger}(0))^{h+1}\Psi^{k}(0)\prod_{j=1}^{N+k}\mathcal{B}(\lambda_j)|0\rangle=& \nonumber\\ \qquad 
\fl =i\sqrt{c}\langle 0|\prod_{j=1\atop j\neq m}^{N+h+1}\mathcal{C}(\mu_j)(\Psi^{\dagger}(0))^{h}\Psi^{k}(0)\prod_{j=1}^{N+k}\mathcal{B}(\lambda_j)|0 \rangle&\nonumber \\ \qquad 
\fl +(i\sqrt{c})^2h\sum_{t=1\atop t\neq m}^{N+h+1}\prod_{o=1\atop {o\neq m\atop o\neq t}}^{N+h+1}f(\mu_o,\mu_t)\langle 0|\prod_{j=1\atop {j\neq m\atop j\neq t}}^{N+h+1}\mathcal{C}(\mu_j)(\Psi^{\dagger}(0))^{h-1}\Psi^{k}(0)\prod_{j=1}^{N+k}\mathcal{B}(\lambda_j)|0\rangle  .
\label{a7}
\end{eqnarray}
Using now (\ref{field_action_2}) we have
\begin{eqnarray}
\fl (i\sqrt{c})\sum_{t=1\atop t\neq m}^{N+h+1}\prod_{o=1\atop {o\neq m\atop o\neq t}}^{N+h+1}f(\mu_o,\mu_t)\langle 0|\prod_{j=1\atop {j\neq m\atop j\neq t}}^{N+h+1}\mathcal{C}(\mu_j)(\Psi^{\dagger}(0))^{h-1}\Psi^{k}(0)\prod_{j=1}^{N+k}\mathcal{B}(\lambda_j)|0\rangle= &\nonumber \\ \qquad \quad
\fl =\langle 0|\prod_{j=1\atop j\neq m}^{N+h+1}\mathcal{C}(\mu_j)(\Psi^{\dagger}(0))^{h}\Psi^{k}(0)\prod_{j=1}^{N+k}\mathcal{B}(\lambda_j)|0\rangle\ ,&
\label{a8}
\end{eqnarray}
showing that (\ref{a5}) is true also for $h+1$. 
The proof of (\ref{a5}) by induction is completed and so property $2$ of Prop. \ref{properties} is true.

Property $3$ is known from the theory of ABA \cite{korepin, norms} so we finally discuss property $4$ of Prop. \ref{properties}. The general case for $h, k\geq 0$ can be proven once again by induction. The case $h=k=0$, corresponding to the scalar product of Bethe states, is well known from the general theory of ABA \cite{korepin, norms, slavnov_89}. Using the action (\ref{field_action}), (\ref{field_action_2}) of the local fields on Bethe states one can proceed by induction as done before for property $2$. A detailed derivation is also given in Appendix B of Ref. \cite{pozsgay_2}. Since it presents no further difficulty, we will not report a derivation here and we refer to \cite{pozsgay_2} for the details.

\section{Proof of Proposition \ref{uniqueness}} \label{proof_uniqueness}
We present here a proof of Prop. \ref{uniqueness}, following \cite{slavnov_89}.

Suppose that the function $\mathcal{H}_{N,M}(\{\mu_j\}_{j=1}^{N},\{\lambda_j\}_{j=1}^{M})$ satisfies properties $1-4$ of Prop. \ref{properties} for a specific value of $(h,k)$ and for every $N\geq h$, $M\geq k$. 
We suppose without loss of generality that $k\geq h$ (the proof is analogous in the case $k\leq h$) and define $\ell=k-h$. Note that  by hypothesis we have $M=N+\ell$. Consider then quantity
\begin{eqnarray}
\fl\Gamma_{N}(\{\mu_j\}_{j=1}^{N},\{\lambda_j\}_{j=1}^{N+\ell})=&\nonumber\\
=\mathcal{H}_{N,N+\ell}(\{\mu_j\}_{j=1}^{N},\{\lambda_j\}_{j=1}^{N+\ell})-\mathcal{F}^{h,k}_{N,N+\ell}(\{\mu_j\}_{j=1}^{N},\{\lambda_j\}_{j=1}^{N+\ell})\ .&
\label{quantity}
\end{eqnarray}
We prove by induction on $N$ that
\begin{equation}
\Gamma_{N}(\{\mu_j\}_{j=1}^{N},\{\lambda_j\}_{j=1}^{N+\ell})=0\ ,\qquad N\geq h .
\label{to_prove_2}
\end{equation}
This is enough to prove Prop. \ref{uniqueness}. The base of the induction is $N=h$ and it is true since by hypothesis $\mathcal{H}_{N,M}$ satisfies property $1$ of Prop. \ref{properties}.

Suppose that (\ref{to_prove_2}) is true for a given $N\geq h$. We show that it is also true for $N+1$. To see this, consider $\Gamma_{N+1}$ as a function of $\mu_1$ . Thanks to property $3$ it is a rational function with at most first order poles at $\lambda_j$, $j=1,\ldots, N+1+\ell$. Now, because both $\mathcal{H}_{N+1,N+1+\ell}$ and $\mathcal{F}_{N+1,N+1+\ell}$ satisfy property $4$, we can compute the residue at $\mu_1=\lambda_k$
\begin{eqnarray}
\fl \Gamma_{N+1}(\{\mu_j\},\{\lambda_j\})\Big|_{\mu_1\to\lambda_k}&\propto& \mathcal{H}_{N,N+\ell}(\{\mu_j\}_{j\neq 1},\{\lambda_j\}_{j\neq k})-\mathcal{F}^{h,k}_{N,N+\ell}(\{\mu_j\}_{j\neq 1},\{\lambda_j\}_{j\neq k}))\nonumber\\ 
\fl &=&\Gamma_{N}(\{\mu_j\}_{j\neq 1},\{\lambda_j\}_{j\neq k})=0\ ,
\end{eqnarray}
where in the last line we used the inductive hypothesis. Thus $\Gamma_{N+1}$ as a function of $\mu_1$ does not have poles. Furthermore, since $\mathcal{H}_{N+1,N+1+\ell}$ and $\mathcal{F}_{N+1,N+1+\ell}$ both satisfy property $2$ of Prop. \ref{properties}, they have the same limit as $\mu_1\to\infty$, that is
\begin{equation}
\lim_{\mu_1\to\infty}\Gamma_{N+1}(\{\mu_j\},\{\lambda_j\})=0\ .
\end{equation}
Summarizing, $\Gamma_{N+1}(\{\mu_j\},\{\lambda_j\})$ as a function of $\mu_1$ is a rational function with no poles on the complex plane and vanishing at $\infty$. Invoking Liouville theorem in complex analysis we thus conclude that it is identically $0$ and thus Prop. \ref{uniqueness} is proved.

\section{Form factors in the attractive case}\label{attractive_calculations}
In this appendix we explicitly derive the formulas for the form factors in the attractive regime when one of the Bethe states corresponds to a $N$-string, as for the ground state. Our derivation follows the one presented in Ref. \cite{calabrese_1}, with only few technical complications arising. We discuss in detail the form factor of $\Psi^{R}(0)$, the case of $(\Psi^{\dagger}(0))^2\Psi^2(0)$ being completely analogous.

Define $M=N+R$, and consider the set of rapidities $\{\lambda_j\}_{j=1}^{M}$ arranged as
\begin{equation}
\lambda_j=\lambda+i\frac{\overline{c}}{2}(M+1-2j)+i\delta_{j}\ .
\label{gs}
\end{equation}
As a first ingredient we need the following expansion for $\{\delta_{j}\}\to 0$
\begin{equation}
K(\lambda_j,\lambda_k)=\left\{
	\begin{array}{cc}
		\frac{1}{\delta_{j,j+1}}+\mathcal{O}(1)\ , & k=j+1\ ,\\
		\frac{1}{\delta_{j-1,j}}+\mathcal{O}(1)\ , & k=j-1\ ,\\
		\mathcal{O}(1)\ , & k\neq j\pm 1\ ,
	\end{array}
	\right.
\label{k_attractive}
\end{equation}
where we defined $\delta_{j,k}=\delta_{j}-\delta_{k}$. We define further
\begin{equation}
V_{j}=\frac{\prod_{m=1}^{N}(\mu_{m}-\lambda_j)}{\prod_{m=1\atop m\neq j}^{M}(\lambda_m-\lambda_j)}\ , \qquad j=0,1,\ldots, M+1\ ,
\label{vv}
\end{equation}
where $\lambda_0$ and $\lambda_{M+1}$ are given by equation (\ref{gs}). Finally, we need the following expansions, which can be easily derived
\begin{equation}
		\widetilde{V}^{+}_{R,j}=\left\{
	\begin{array}{cc}
		V_0\ , & j=1\ ,\\
		\frac{-i}{\delta_{j-1, j}}V_{j-1}\ , & j=2,\ldots, M\ ,
	\end{array}
	\right.
\label{exp1}
\end{equation}
\begin{equation}
	\widetilde{V}^{-}_{R,j}=\left\{
	\begin{array}{cc}
		\frac{i}{\delta_{j,j+1}}V_{j+1}\ , & j=1, \ldots , M-1\ ,\\
		V_{M+1}\ , & j=M\ ,
	\end{array}
	\right.
\label{exp2}
\end{equation}
where $\widetilde{V}^{\pm}_{R,j}$ is given in (\ref{wvpm}).

We now plug the rapidities (\ref{gs}) into (\ref{general_ff}) and take the limit $\{\delta_j\}\to 0$. Consider the matrix 
\begin{equation}
\fl W_{jk}=\delta_{jk}\frac{\widetilde{V}^{+}_{R,j}-\widetilde{V}^{-}_{R,j}}{i}+\frac{\prod_{m=1}^N(\mu_{m}-\lambda_j)}{\prod_{m=1\atop m\neq j}^{N+R}(\lambda_m-\lambda_j)}\left[K(\lambda_j,\lambda_k)-K(\lambda_p,\lambda_k)K(\lambda_s,\lambda_j)\right]\ ,
\label{w_matrix}
\end{equation}
so that
\begin{equation}
\prod_{j=1}^{M}\left(\widetilde{V}^{+}_{R,j}-\widetilde{V}^{-}_{R,j}\right)\mathrm{det\ }_{M}(\delta_{jk}+\widetilde{U}^{(R)}_{jk})=i^{M}\mathrm{det\ }_{M}W_{jk}\ .
\label{aux_2}
\end{equation}
The matrix $W_{jk}$ has singular behavior in the limit $\{\delta_j\}\to 0$. Since $\lambda_p$ is an arbitrary parameter we can choose $\lambda_p=\lambda_M$; with this choice the leading order of the matrix $W_{jk}$ is
\begin{equation*}
\fl \left(
\begin{array}{ccccccc}
\frac{-V_2}{\delta_{1,2}} & \frac{V_1}{\delta_{1,2}} & 0 &\ldots & 0 & \frac{-\mathcal{W}_1}{\delta_{M-1,M}}&0\\
\frac{V_2}{\delta_{1,2}} & -\frac{V_1}{\delta_{1,2}}-\frac{V_3}{\delta_{2,3}} & \frac{V_2}{\delta_{2,3}} &\ldots & 0 & \frac{-\mathcal{W}_2}{\delta_{M-1,M}}&0\\
\vdots &\vdots &\vdots & \ddots & \vdots &\vdots & \vdots\\
0&0&0&\cdots&  \frac{-V_{M-3}}{\delta_{M-4,M-3}}-\frac{V_{M-1}}{\delta_{M-2,M-1}}& \frac{V_{M-2}}{\delta_{M-2,M-1}}-\frac{\mathcal{W}_{M-2}}{\delta_{M-1,M}}& 0\\
0&0&0&\cdots& \frac{V_{M-1}}{\delta_{M-2,M-1}}& \frac{-V_{M-2}}{\delta_{M-2,M-1}}-\frac{\mathcal{W}_{M-1}+V_M}{\delta_{M-1,M}}& \frac{V_{M-1}}{\delta_{M-1,M}} \\
0&0&0&\cdots& 0& \frac{V_{M}}{\delta_{M-1,M}}-\frac{\mathcal{W}_{M}}{\delta_{M-1,M}}& -\frac{V_{M-1}}{\delta_{M-1,M}} \\
\end{array}
\right)
\end{equation*}
where we defined $\mathcal{W}_j=V_j K(\lambda_s,\lambda_j)$. The determinant of this matrix can be computed as follows. First, add the first row to the second, then the second to the third, and iterate this procedure until the last row. The last row of the resulting matrix has only one entry different from $0$. Using Laplace expansion on the last row and the fact that the determinant of a triangular matrix is the product of its diagonal elements, we finally have
\begin{equation}
\fl\mathrm{det}_{M}W_{jk}=(-1)^{M-2}\frac{V_{M-1}}{\delta_{M-1,M}}\left[\prod_{j=1}^{M-1}\frac{1}{\delta_{j,j+1}}\right]\left[\prod_{j=2}^{M-1}V_{j}\right]\left[\sum_{j=1}^{M}K(\lambda_s,\lambda_j)V_{j}\right]\ .
\label{final_det}
\end{equation}
Using (\ref{aux_2}) we can plug (\ref{final_det}) into (\ref{general_ff}). One then uses (\ref{identity_1}) of appendix \ref{useful_identities} and the following identity, which is easily derived for rapidities satisfying (\ref{gs})
\begin{equation}
\prod_{m=1\atop m\neq j}^{M}\left(\lambda_m-\lambda_j\right)=(i\overline{c})^{M-1}(-1)^{M-j}(j-1)!(M-j)!\ .
\end{equation}
Putting everything together, after simple manipulations we get (\ref{attractive_field}).

\section{Useful formulas}\label{useful_identities}
In this appendix we discuss identities involving sums of rational functions. These are useful when studying the properties of the determinant of matrices like those appearing in (\ref{rho_squared}) and (\ref{general_ff}). The prototypical example is
\begin{equation}
i\sum_{j=1}^{N+R}K(\lambda_s,\lambda_j)\frac{\prod_{m=1}^{N}(\mu_m-\lambda_j)}{\prod_{m\neq j}^{N+R}(\lambda_m-\lambda_j)}=-\left(\widetilde{V}_{R,s}^+-\widetilde{V}_{R,s}^-\right)\ ,
\label{identity_1}
\end{equation}
where $K(\lambda,\mu)$ and $\widetilde{V}_{R,s}^{\pm}$ are given in (\ref{k_function}), (\ref{wvpm}) respectively. Identity (\ref{identity_1}) is obtained using the residues theorem on the complex function
\begin{equation}
g_s(z)=\frac{1}{(z-\lambda_s-ic)(z-\lambda_s+ic)}\frac{\prod_{m=1}^N(\mu_m-z)}{\prod_{m=1}^{N+R}(\lambda_m-z)}\ .
\label{complex_g}
\end{equation}
Indeed the function $g_s(z)$ has first order poles for $z=\lambda_j$, $j=1,\ldots, \lambda_{N+R}$ and for $z=\lambda_{s}\pm ic$ while it is easy to see that it has vanishing residue at infinity. Using the fact that the sum of the residues has to be zero one immediately arrives at identity (\ref{identity_1}).

\section{Derivation of equivalent formulas}\label{derivation_equivalent}
In this appendix we show how to derive formulas (\ref{different_momentum}) and (\ref{equal_momentum}) from (\ref{rho_squared}). First define the matrix $E^{\ell}$ as 
\begin{equation}
E^{\ell}_{jk}=\left\{
	\begin{array}{cc}
	\frac{K(\lambda_p,\lambda_k)}{V^+_j-V^{-}_j}\ , & \mathrm{if\ } j=\ell\ ,\\
	\delta_{jk}+\frac{i}{V^+_j-V^{-}_j}\frac{\prod_{m=1}^N(\mu_{m}-\lambda_j)}{\prod_{m=1\atop m\neq j}^{N}(\lambda_m-\lambda_j)}K(\lambda_j,\lambda_k)\ , &\mathrm{otherwise\ .}
	\end{array}
	\right.
\label{e_matrix}
\end{equation}
Consider the matrix $\delta_{jk}+U_{jk}$ appearing in (\ref{rho_squared}). For $j=1,\ldots,N$, $j\neq \ell$, add to row $\ell$ row $j$ multiplied by $\Theta_{j}/\Theta_{\ell}$, with $\Theta_{j}$ defined as
\begin{equation}
\Theta_{j}=V^{+}_j-V^{-}_j\ ,
\label{xi_0}
\end{equation}
and where $V^{\pm}_j$ is given in (\ref{vpm}).
With this procedure one obtains
\begin{equation}
\mathrm{det}_{N}(\delta_{jk}+U_{jk})=(V^{+}_s-V^{-}_s)\mathrm{det}_{N}M^{\ell}\ ,
\label{e1}
\end{equation}
where the matrix $M^{\ell}$ is defined by
\begin{equation}
M^{\ell}_{jk}=\left\{
	\begin{array}{cc}
	\frac{K(\lambda_p,\lambda_k)}{V^+_j-V^{-}_j}\ , & \mathrm{if\ } j=\ell\ ,\\
	\delta_{jk}+U_{jk}\ , &\mathrm{otherwise\ }.
	\end{array}
	\right.
\label{m_matrix}
\end{equation}
For $j=1,\ldots, N$, $j\neq \ell$, add now to row $j$ row $\ell$ multiplied by $i K(\lambda_s,\lambda_j)\Xi_{j}$, where
\begin{equation}
\Xi_{j}=\frac{\prod_{m=1}^{N}(\mu_{m}-\lambda_j)}{\prod_{m=1\atop m\neq j}^{N}(\lambda_{m}-\lambda_j)}\ .
\label{aux}
\end{equation}
Using the above procedure we see that $\mathrm{det}_{N}M^{\ell}=\mathrm{det}_{N}E^{\ell}$, and thus we obtain
\begin{equation}
\mathrm{det}_{N}(\delta_{jk}+U_{jk})=(V^{+}_s-V^{-}_s)\mathrm{det}_{N}E^{\ell}\ .
\label{e2}
\end{equation}
Now consider the matrix $\delta_{jk}+U^{(1)}_{jk}$ in equation (\ref{different_momentum}). For $j=1,\ldots, N$, $j\neq \ell$ add to row $\ell$ row $j$ multiplied by $\Theta_{j}/\Theta_{\ell}$, with $\Theta_{j}$ is defined in (\ref{xi_0}). Using the the identity
\begin{equation}
\sum_{j=1}^N\frac{\prod_{m=1}^N(\mu_m-\lambda_j)}{\prod_{m\neq j}^N(\lambda_m-\lambda_j)}=\sum_{j=1}^{N}(\mu_j-\lambda_j)\ ,
\label{identity_momentum}
\end{equation}
which can be derived with the techniques described in  \ref{useful_identities}, one obtains 
\begin{equation}
\mathrm{det}_{N}(\delta_{jk}+U^{(1)}_{jk})=-i(P_{\mu}-P_{\lambda})\mathrm{det}_{N}\widetilde{M}^{\ell}\ ,
\label{e3}
\end{equation}
where the matrix $\widetilde{M}^{\ell}$ is defined as
\begin{equation}
\widetilde{M}^{\ell}_{jk}=\left\{
	\begin{array}{cc}
	\frac{K(\lambda_p,\lambda_k)}{V^+_j-V^{-}_j}\ , & \mathrm{if\ } j=\ell\ ,\\
	\delta_{jk}+U^{(1)}_{jk}\ ,	&\mathrm{otherwise}.
	\end{array}
	\right.
\label{wm_matrix}
\end{equation}
Using again simple manipulations and the multilinearity of the determinant one sees that
\begin{equation}
\mathrm{det}_{N}\widetilde{M}^{\ell}=\mathrm{det}_{N}E^{\ell}\ .
\label{e4}
\end{equation}
Putting together (\ref{e2}), (\ref{e3}) and (\ref{e4}) it is now straightforward to see that (\ref{different_momentum}) is indeed equivalent to (\ref{rho_squared}).

We now derive (\ref{equal_momentum}) from (\ref{rho_squared}). From (\ref{e2}) we see that $\mathrm{det}_NE^{\ell}=\mathrm{det}_NE^{m}$ for $\ell\neq m$, so using (\ref{e4}) we can write
\begin{equation}
\mathrm{det}_{N}(\delta_{jk}+U_{jk})=(V^{+}_s-V^{-}_s)\frac{1}{N}\sum_{\ell=1}^{N}\mathrm{det}_{N}(\widetilde{M}^{\ell}_{jk})\ .
\label{e5}
\end{equation}
The sum of determinants in the r.h.s of (\ref{e5}) can be rewritten as follows
\begin{equation}
\sum_{\ell=1}^{N}\mathrm{det}_{N}(\widetilde{M}^{\ell}_{jk})=-i\left(\mathrm{det}_{N}(\delta_{jk}+U^{(2)}_{jk})-\mathrm{det}_N(\delta_{jk}+U^{(1)}_{jk})\right)\ ,
\label{e6}
\end{equation}
where $U^{(1)}_{jk}$ and $U^{(2)}_{jk}$ are given in (\ref{u_1}), (\ref{u_2}). The validity of Eq. (\ref{e6}) follows from 
$U^{(2)}_{jk}=U^{(1)}_{jk}+iK(\lambda_p,\lambda_k)/(V^{+}_{j}-V^{-}_j)$ 
and from the fact that $N_{jk}=iK(\lambda_p,\lambda_k)/(V^{+}_{j}-V^{-}_j)$ is a rank $1$ matrix.

Now, from (\ref{e3}) we see that $\mathrm{det}_{N}(\delta_{jk}+U^{(1)}_{jk})\propto(P_{\lambda}-P_{\mu})$, and thus its contribution vanishes when we consider Bethe states having the same momentum. Putting together this observation with (\ref{e5}) and (\ref{e6}) it is then immediate to obtain (\ref{equal_momentum}) from (\ref{rho_squared}).

\end{document}